\documentclass[aps,amsmath,amssymb,twocolumn,floatfix,author-year]{revtex4-2}
\usepackage{graphicx}
\usepackage{amsmath}
\usepackage{dcolumn}
\usepackage{bm}

\AtBeginDocument{%
  \setlength{\abovedisplayskip}{6pt}%
  \setlength{\belowdisplayskip}{6pt}%
  \setlength{\abovedisplayshortskip}{4pt}%
  \setlength{\belowdisplayshortskip}{4pt}%
  \setlength{\textfloatsep}{12pt}%
  \setlength{\floatsep}{10pt}%
  \setlength{\intextsep}{10pt}%
  \setlength{\abovecaptionskip}{6pt}%
}
\makeatletter
\def\bibsection{%
  \par
  \addvspace{19\p@}%
  \noindent\rule{\columnwidth}{0.5\p@}\par
  \nobreak\@nobreaktrue
  \addvspace{6\p@}%
  \par
}
\makeatother
\begin{document}
\preprint{APS/123-QED}
\title{Nonequilibrium Dynamics of R-loop Formation in the CRISPR-Cas9 System}
\author{Yu Liu}
\author{Bing Miao}
 \email{bmiao@ucas.ac.cn}
\affiliation{
Center of Materials Science and Optoelectronics Engineering, College of Materials Science and Opto-Electronic Technology, University of Chinese Academy of Sciences (UCAS), Beijing 100049, China.
}

\begin{abstract}
The binding specificity of CRISPR-Cas9 is governed by the stepwise formation of an R-loop, in which the sgRNA displaces the nontarget DNA strand and hybridizes with the target strand in discrete base pair steps. Here we develop a nonequilibrium dynamics framework that models R-loop formation as a continuous-time Markov birth-death process over twenty base pairs. Each transition rate is derived from Eyring transition-state theory and the Bell model, and combines three physical contributions, i.e., the torsional strain generated by the B-form to A-form helical conversion at the unwinding front, a base stacking correction that couples neighboring base pairs, and the allosteric work supplied by Cas9 conformational changes. We further derive the entropy production of the process, which is decomposed into activation, stacking, torsional, and allosteric terms. The resulting prediction model uses no external thermodynamic database and reaches Spearman correlations of $0.84$ and $0.82$ in two independent libraries of dCas9-sgRNA mismatch variants measured in vivo. The mismatch-type dependence is carried by one twist deviation and one pairing weight per type. Our model recovers the seed region sensitivity, the PAM-proximal activity gradient, and the moderate cost of dDrG mismatches, while isolating the high-activity double mismatch tail as the regime in which the present free-energy mapping is inadequate. The framework gives quantitative predictions on how DNA supercoiling and sequence context modulate binding affinity, and provides a mechanistic basis for sgRNA design.
\end{abstract}

\maketitle
\section{Introduction}
\label{sec:introduction}

The CRISPR-Cas9 system, in which a single-guide RNA (sgRNA) directs cleavage of a complementary DNA sequence~\cite{Jinek2012}, has become a general tool for genome editing and regulation~\cite{Doudna2014}. Its specificity is nevertheless incomplete, because the sgRNA tolerates mismatches and Cas9 can therefore bind or cleave at unintended genomic sites~\cite{Fu2013,Hsu2013}. This off-target activity remains the central obstacle to therapeutic use, and controlling it requires a quantitative physical description of how mismatches modulate Cas9 binding and cleavage.

The binding specificity of Cas9 is governed by the stepwise formation of an R-loop. In this process the sgRNA displaces the nontarget DNA strand and hybridizes with the target strand, and the resulting RNA--DNA heteroduplex is accommodated in a channel of the protein~\cite{Anders2014,Nishimasu2014}. The displaced nontarget strand is held in a groove on the protein surface, so the guide-target match is read base by base along the heteroduplex, which adopts an A-form geometry in contrast to the flanking B-form DNA. Single-molecule experiments show that the R-loop does not form in one concerted event. It grows directionally from the PAM-proximal end toward the distal end in discrete base pair steps, and DNA unwinding, rather than sgRNA--DNA hybridization, is the rate-limiting step~\cite{Sternberg2014,Szczelkun2014,Gong2018}. These steps are sensitive to guide-target mismatches and to DNA supercoiling, so torsional stress is mechanically coupled to R-loop propagation~\cite{Ivanov2020}. The seed region, spanning about ten base pairs from the PAM-proximal end, sets the specificity of CRISPR interference~\cite{Semenova2011}, and high-throughput binding assays have mapped the sequence determinants of off-target binding and unbinding~\cite{Boyle2017}. The mismatch tolerance profiles for cleavage and for binding are qualitatively different, which implies that cleavage cannot be inferred from binding affinity alone and that binding and cleavage are distinct stages of the same sequential process. The stepwise, reversible, and mechanically driven nature of R-loop formation calls for a master equation description, in which the probability distribution over the number of unwound base pairs evolves through a continuous-time Markov birth-death process.

CRISPR-Cas9 targeting has been theoretically studied along two lines. The first describes guide-target hybridization in kinetic terms: a model of hybrid formation kinetics reproduces the off-targeting rules of Cas nucleases~\cite{Klein2018}, and its extension predicts SpCas9 binding and cleavage activity and classifies off-targets better than sequence-based scores~\cite{EslamiMossallam2022}, while the reversible Markov chain fitted by Feng~\textit{et~al.} belongs to the same family~\cite{Feng2021}. The second assigns each mismatch an equilibrium free energy, as in the duplex energy model used to score off-targets~\cite{Alkan2018}. In both routes specificity is read from equilibrium occupancies of the hybrid, and the equilibrium route is further limited by the small number of mismatch types with tabulated parameters.

Discrimination itself, however, is not an equilibrium property. Cas9 proofreads the R-loop through a conformational checkpoint that commits the enzyme to cleavage only when full complementarity is sensed~\cite{Chen2017}, so mismatched targets are rejected by kinetics rather than by binding affinity. Kinetic proofreading explains how such a checkpoint improves accuracy, but only at a price, since the extra discrimination requires an irreversible step driven by chemical energy~\cite{Hopfield1974}. The thermodynamics of error correction makes the price explicit, bounding the fidelity of a discriminator by the entropy production it must sustain~\cite{SartoriPigolotti2015}, and stochastic thermodynamics supplies the tools to compute that cost for a given reaction network~\cite{Seifert2012,Gaspard2004,Rao2016}.

The entropy production of R-loop formation has nevertheless remained unquantified. Here we derive it for the sequential R-loop process and decompose it into activation, stacking, torsional, and allosteric contributions, which links the fidelity of Cas9 to the dissipation along its reaction pathway.

We test this framework against the binding data of Feng~\textit{et~al.}~\cite{Feng2021}, who measured over 3400 saturated single and double mismatch variants of dCas9-sgRNA in two independent libraries in living bacterial cells. Our starting point is that R-loop formation is a sequential process rather than an equilibrium binding problem, because the guide invades the duplex one base pair at a time from the PAM-proximal end. We therefore model it as a continuous-time Markov birth-death process on the state space $\{0,1,\ldots,20\}$, where the state counts the consecutively unwound base pairs, and each transition involves a single base pair and comprises duplex unwinding followed by sgRNA-target strand binding. The rates follow from Eyring transition-state theory~\cite{Eyring1935} and the Bell model~\cite{Bell1978}, and combine three physical contributions, i.e., the torsional strain of the B-form to A-form conversion at the unwinding front, a base stacking correction that couples neighboring base pairs, and the allosteric work supplied by Cas9. From the same process we derive the exact entropy production of a trajectory, which splits into a medium part from the twenty transitions and a system part from the boundary states.

The resulting physically constrained model uses no external thermodynamic database. It reaches $R^2=0.68\pm0.06$ and $0.64\pm0.05$, with Spearman correlations of $0.84$ and $0.82$, over seven independent partitions. It recovers the seed-region sensitivity, the PAM-proximal activity gradient, and the moderate cost of dDrG mismatches, with the mismatch-type dependence carried by one twist deviation and one pairing weight per type, and it isolates the high-activity double mismatch tail as the regime in which the present free-energy mapping is inadequate.

The framework links the specificity of Cas9 to a small set of physical quantities. The torsional and stacking parameters give quantitative predictions of how supercoiling and sequence context shift binding affinity, and the parameter structure can be retrained on new libraries without a complete mismatch database, which offers a mechanistic route to improve existing sgRNA design tools~\cite{Doench2016,MorenoMateos2015}.

The paper is organized as follows. Sec.~\ref{sec:model} formulates the sequential model of R-loop formation and the transition rates it implies. Sec.~\ref{sec:single_bp} analyzes the single base pair dynamics. Sec.~\ref{sec:thermo} constructs the thermodynamic model of the formation efficiency. Sec.~\ref{sec:entropy} derives the entropy production and its decomposition. Sec.~\ref{sec:physics_model} validates the physically constrained prediction model on the two libraries. Finally, in Sec.~\ref{sec:conclusion} we present our conclusions. Detailed derivations are given in Appendices~\ref{app:A} and~\ref{app:B}.

\section{Sequential model of R-loop formation}
\label{sec:model}

\begin{figure}[!htbp]
	\centering
	\includegraphics[width=\linewidth]{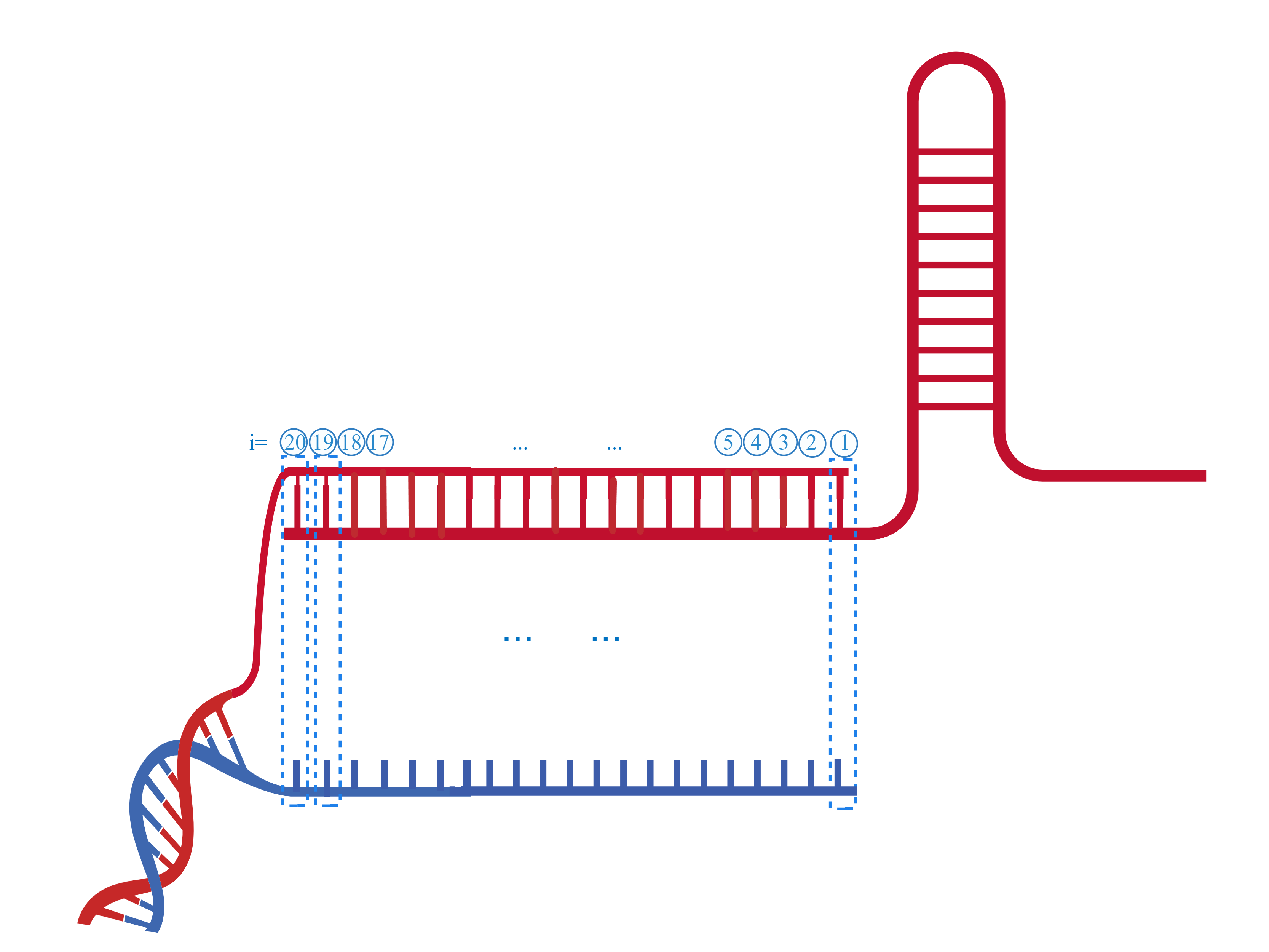}
	\caption{Schematic of the base pairing along the 20-bp protospacer. Circled numbers label positions along the protospacer, each joined by a leader line to the base pair it marks. The duplex continues beyond the protospacer on the right. R-loop formation proceeds sequentially in $20$ base pair steps, indexed $i=1,\ldots,20$ from the PAM-proximal to the PAM-distal end.}
	\label{fig1}
\end{figure}

R-loop formation proceeds as a sequential unwinding of the DNA duplex that starts at the PAM-proximal end and advances base pair by base pair toward the distal end. Because Cas9 recognizes a 20-bp protospacer, complete R-loop formation comprises 20 sequential steps, indexed $i=1,2,\ldots,20$ from the PAM-proximal to the PAM-distal end, as schematically illustrated in Fig.~\ref{fig1}. Each step corresponds to the transition of one base pair from the closed to the unwound state and comprises two strictly ordered microscopic subprocesses, DNA unwinding and sgRNA binding. The DNA duplex at position~$i$ can unwind only after position~$i{-}1$ has unwound, and the sgRNA can pair with the target strand at position~$i$ only after that position has been opened. As no skipping of intermediate steps has been observed experimentally, we exclude both the direct opening of a distal base pair and the initiation of a distal step before the proximal ones are complete.

Based on this picture, we define the state variable $i(t)$ as the number of consecutively unwound base pairs at time~$t$. Here consecutive means that base pairs~1 through~$i$, counted from the PAM-proximal end, are all unwound while base pair~$i{+}1$ remains closed. The variable takes values $0,1,\ldots,20$, where $i=0$ is the initial state with all 20~base pairs closed and $i=20$ the final state with the R-loop fully formed.

Let $Q_i(t)$ denote the probability that the system is in state~$i$ at time~$t$. Since the system must occupy exactly one state at a given instant, the normalization condition holds automatically:
\begin{equation}
	\sum_{i=0}^{20} Q_i(t) = 1, \quad \forall\, t \geq 0.
\end{equation}
At the initial moment when the Cas9-sgRNA complex first binds to the PAM-containing target DNA, no base pairs have yet unwound. Hence
\begin{equation}
	Q_0(0) = 1, \quad Q_i(0) = 0 \quad (i \geq 1).
\end{equation}

The process is governed by two fundamental physical constraints. First, the state space is ordered. The system must pass through states $0,1,2,\ldots,20$ in sequence, with no skipping. From state~$i$, only transitions to $i{+}1$ or $i{-}1$ are allowed. Direct jumps to $i\pm 2$ or beyond are forbidden, so the process has local dynamics. Second, a local interaction constraint requires that at most one base pair changes state at any given time. Each forward or backward transition involves exactly one step. Together, these constraints define the system as a continuous-time Markov birth-death process on the state space.

Within the birth-death framework, the evolution of $Q_i(t)$ is governed by the master equation.
\begin{equation} \label{eq:master}
	\begin{split}
		\frac{dQ_0}{dt} &= -\lambda_{1,0}\,Q_0 + \lambda_{0,1}\,Q_1, \\
		\frac{dQ_i}{dt} &= \lambda_{i,i-1}\,Q_{i-1}
		- (\lambda_{i-1,i} + \lambda_{i+1,i})\,Q_i \\
		&\quad + \lambda_{i,i+1}\,Q_{i+1}, \qquad (1 \leq i \leq 19), \\
		\frac{dQ_{20}}{dt} &= \lambda_{20,19}\,Q_{19}
		- \lambda_{19,20}\,Q_{20}.
	\end{split}
\end{equation}

 Here $\lambda_{i,j}$ denotes the element of the transition-rate matrix for the transition from state~$j$ to state~$i$. Probability conservation and the steady-state recursive structure follow directly from the mathematical form of Eq.~(\ref{eq:master}), independent of the specific values of $\lambda_{i,i-1}$ and $\lambda_{i-1,i}$. The final state $i=20$ represents the completed R-loop. It is not strictly absorbing, because the reverse transition $20\to19$ is retained in Eq.~(\ref{eq:master}). Cleavage enters as the factor $1-\mu_{\mathrm{cleavage}}$ that multiplies the rate of that reverse transition, so that a complex that has been cleaved can no longer rewind, with $\mu_{\mathrm{cleavage}}$ defined below.

\section{Single base pair dynamics}
\label{sec:single_bp}

In the master equation framework, the forward rate $\lambda_{i,i-1}$ is activated only when the system resides in state~$i{-}1$, which means the first $i{-}1$ base pairs have already unwound. The natural formulation of single base pair dynamics is therefore based on a conditional probability equation, with the condition that all preceding steps are complete. Across the 20 reaction sites, each site obeys the same single-step kinetic scheme. The DNA unwinding and sgRNA-binding subprocesses at step~$i$ can be written as the chemical reactions
\begin{align}
	\text{T-DNA} + \text{NT-DNA}
	&\overset{k^{i}_{\mathrm{on,1}}}{\underset{k^{i}_{\mathrm{off,1}}}{\rightleftharpoons}}
	\text{dDNA}, \label{eq:rxn1} \\
	\text{sgRNA} + \text{T-DNA}
	&\overset{k^{i}_{\mathrm{on,2}}}{\underset{k^{i}_{\mathrm{off,2}}}{\rightleftharpoons}}
	\text{RNA--DNA}, \label{eq:rxn2}
\end{align}
where dDNA denotes the intact double-stranded DNA, RNA--DNA the RNA--DNA hybrid duplex, T-DNA the target DNA strand, and NT-DNA the nontarget DNA strand. The rate constants $k^{i}_{\mathrm{on,1}}$ and $k^{i}_{\mathrm{off,1}}$ govern the closing and unwinding of the DNA duplex at step~$i$, while $k^{i}_{\mathrm{on,2}}$ and $k^{i}_{\mathrm{off,2}}$ govern the binding and dissociation of sgRNA to the exposed target strand. The subscripts~1 and~2 distinguish the two subprocesses.

Because these reactions involve the motion of complementary segments on the same DNA duplex or RNA--DNA hybrid, they are intramolecular processes. The rate equation for either subprocess
takes the unified form
\begin{equation} \label{eq:rate_general}
	\frac{d[C]}{dt} = k^{i}_{\mathrm{on}}\,[O]
	- k^{i}_{\mathrm{off}}\,[C],
\end{equation}
where $[O]$ and $[C]$ denote the fractions of intramolecular open and closed sites, respectively. Consequently, $k^{i}_{\mathrm{on}}$ and $k^{i}_{\mathrm{off}}$ carry dimensions of $\mathrm{s}^{-1}$.

We now derive the evolution equation for the unwinding probability from the above reaction scheme. Define $q_i(t)$ as the conditional probability that the $i$-th base pair is unwound at time~$t$, given that the first $i{-}1$ base pairs have already unwound:
\begin{equation} \label{eq:Si_def}
	q_i(t) = \frac{P_1^{i}(t)}{P_1^{i-1}(t)},
\end{equation}
where $P_1^{i}(t)$ is the probability that at least the first $i$~base pairs are unwound at time~$t$. The master equation provides the probability distribution $Q_m(t)$, the probability of being in the state with exactly $m$~unwound base pairs. By contrast, $P_1^{i}$ is the cumulative probability that the first $i$~base pairs are all unwound, regardless of the states of base pairs $i{+}1$ and beyond. These two descriptions are related by the identity
\begin{equation}
	P_1^{i}(t) = \sum_{m=i}^{20} Q_m(t). \label{eq:P1Q}
\end{equation}

The physical basis for this relation is the sequential nature of unwinding. Since the process starts at the PAM-proximal end and proceeds without skipping, having at least $i$~base pairs unwound is equivalent to the system residing in some state $m \geq i$. From Eq.~(\ref{eq:P1Q}) one further obtains
\begin{align}
	Q_i &= P_1^{i} - P_1^{i+1}, \quad (1 \leq i \leq 19),
	\label{eq:QK119} \\
	Q_{20} &= P_1^{20}. \label{eq:Qk20}
\end{align}
In addition, $P_1^{0}$ represents the probability of PAM recognition. Since we focus on R-loop formation after PAM binding, we set $P_1^{0} = 1$.

Combining the physical picture with the reaction scheme, the dynamics of $q_i$ for the DNA-unwinding subprocess reads
\begin{equation} \label{siggmadot}
	\dot{q}_i = k^{i}_{\mathrm{off,1}}\,(1 - q_i) - (1-\mu_{\mathrm{cleavage}}\delta_{i,20}) k^{i}_{\mathrm{on,1}}\,q_i\,(1 - q_{i+1}),
\end{equation}
where $\mu_{\mathrm{cleavage}}$ is the cleavage efficiency of the Cas9 complex, with $0\leq\mu_{\mathrm{cleavage}}<1$, and $\delta_{i,20}$ is the Kronecker delta. The forward term $k^{i}_{\mathrm{off,1}}(1-q_i)$ accounts for the probability $(1-q_i)$ that base pair~$i$ is still closed, given that pairs~1 through $i{-}1$ are already open. This allows base pair~$i$ to unwind. The backward term $k^{i}_{\mathrm{on,1}}q_i(1-q_{i+1})$ represents the sequentiality constraint. Pair~$i$ can rewind only when it is open and pair~$i{+}1$ is closed. Under this constraint, the dynamics of step~$i{-}1$ is much slower than that of step~$i$. When observing step~$i$, the state of the first $i{-}1$ base pairs appears frozen, so that $\dot{P}_1^{i-1} \approx 0$. Eq.~(\ref{siggmadot}) can then be rewritten as
\begin{equation} \label{dotP}
	\frac{\dot{P}_1^{i}}{P_1^{i-1}}
	= k^{i}_{\mathrm{off,1}}\,(1 - q_i)
	-  (1-\mu_{\mathrm{cleavage}}\delta_{i,20})k^{i}_{\mathrm{on,1}}\,q_i\,(1 - q_{i+1}).
\end{equation}

For the sgRNA-binding subprocess, we assume that binding at site~$i$ is triggered only after the target DNA strand at that site has been exposed by unwinding. Define $P_2^{i}(t)$ as the probability that the sgRNA segment involved in step~$i$ is bound to the exposed target strand at time~$t$. The binding probability $P_2^{i}$ obeys
\begin{equation} \label{p2}
	\frac{dP_2^{i}}{dt}
	= k^{i}_{\mathrm{on,2}}\,(1 - P_2^{i})
	- k^{i}_{\mathrm{off,2}}\,P_2^{i},
\end{equation}
where $k^{i}_{\mathrm{on,2}}(1-P_2^{i})$ is the binding rate, proportional to the fraction of unbound sites, and $k^{i}_{\mathrm{off,2}}P_2^{i}$ is the dissociation rate, proportional to the fraction of bound sites.

 Now we express the master equation transition rates $\lambda_{i,i-1}$ and $\lambda_{i-1,i}$ in terms of the microscopic rate constants. Recalling the relations
\begin{align}
	Q_i &= P_1^{i} - P_1^{i+1}, \quad (1 \leq i \leq 19),
	\label{eq:QK119b} \\
	Q_{20} &= P_1^{20}, \label{eq:Qk20b}
\end{align}
a rigorous derivation (see Appendix~A) yields the exact expressions
\begin{align}
	\lambda_{i,i-1} &= k^{i}_{\mathrm{off,1}},
	\label{eq:lambda_full} \\
	\lambda_{i-1,i} &=(1-\mu_{\mathrm{cleavage}}\delta_{i,20}) k^{i}_{\mathrm{on,1}}.
	\label{eq:mu_full}
\end{align}

\section{Thermodynamic model for R-loop formation efficiency}
\label{sec:thermo}

In this section, we connect the dynamics developed above to thermodynamics. In each of the 20 substeps the first microscopic process is the unwinding of the DNA duplex, which includes several contributions: the attractive force of the sgRNA bases, the Cas9 conformational change that promotes unwinding, and one frequently overlooked factor, the torsional strain that builds up where the A-form RNA--DNA hybrid meets the B-form duplex and that must be overcome for each base pair converted from B-form to A-form.

Quantifying this torsional contribution requires a mechanical analysis. Both the sgRNA and the nontarget DNA strand bind to the target DNA strand, creating a boundary on the target strand where one side is occupied by sgRNA and the other by the nontarget strand. On each side of this boundary, the helix adopts a different conformation. The RNA--DNA hybrid formed by sgRNA and the target strand typically assumes an A-form helix, whereas the DNA duplex retains the B-form structure. These two helical forms differ in their twist per base pair, about $32.73^{\circ}$ for A-form and $34.29^{\circ}$ for B-form. One step advances the junction by the rise of a single base pair along the helix axis, $\Delta x\approx0.34$~nm. Consequently, at the boundary on the target strand, the A- and B-form segments exert torques $\tau_A$ and $\tau_B$, respectively, where
\begin{equation}
	\tau_A = C\,\frac{32.73^{\circ}}{\Delta x}, \qquad
	\tau_B = C\,\frac{34.29^{\circ}}{\Delta x},
\end{equation}
with $d\theta/ds \approx \Delta\theta_0/\Delta x$ and $C$ the torsional stiffness. The two helices are held by the protein, so the twist difference between them is accommodated at the junction. We take this residual mismatch to be $\Delta\theta_0\approx1.56^{\circ}$, the difference between the A-form and the B-form twist angles, which is consistent with the near-straight heteroduplex seen in cryo-EM structures of the Cas9-sgRNA-DNA complex~\cite{Anders2014,Nishimasu2014}. If the boundary advances from the B-form to the A-form side at velocity~$v$, the instantaneous power delivered by the torque is
\begin{equation} \label{P}
	P(t) = \tau_{\mathrm{ex}}\,\partial_t\theta
	= (\tau_A - \tau_B)(v\,\partial_s\theta)
	= \frac{v\,(\tau_A - \tau_B)^2}{C}.
\end{equation}
The transition of the boundary across site~$i$ corresponds to the formation of A-form structure at position~$i{-}1$ followed by the unwinding of the DNA duplex at position~$i$, i.e., the B$\to$A conversion. The torsional energy that must be supplied to carry the junction through one step is therefore
\begin{equation} \label{DeltaW}
	\Delta W_0 = \int_0^{\Delta x/v} P(t)\,dt
	= \frac{v\,(\tau_A - \tau_B)^2}{C}\,\frac{\Delta x}{v}
	= \frac{C\,(\Delta\theta_0)^2}{\Delta x}.
\end{equation}

Equation~(\ref{DeltaW}) holds under ideal conditions, and $C/\Delta x$ can be interpreted as an effective local torsional stiffness. The torsional energy of Eq.~(\ref{DeltaW}) is a cost per step: it raises the barrier for opening a base pair at the junction and decreases it for closing, and it enters the rate constants below with that sign. In a perfectly matched hybrid the sgRNA pairs with the target strand in standard Watson--Crick geometry, adenine with thymine and guanine with cytosine. However, if the sgRNA and target DNA sequences are not perfectly complementary, mismatches or even complete noncomplementary sites may arise. In double-stranded nucleic acids, mismatched base pairs typically induce local helical distortions that alter the twist angle between adjacent base pairs. The magnitude and sign of these angular perturbations depend on the mismatch type and the local sequence context. Moreover, the effect of a mismatch is not confined to the
current base pair. It can propagate to one or two neighboring pairs, albeit with diminishing amplitude.

To account for mismatches, we generalize $\Delta W_i$ as
\begin{equation} \label{DeltaWi}
	\Delta W_i = \frac{C\,(\Delta\theta_i)^2}{\Delta x},
\end{equation}
where $\Delta\theta_i$ denotes the twist-angle difference between the two helical forms at site~$i$:
\begin{equation} \label{theta}
	\Delta\theta_i = \Delta\theta_0
	+ \delta\theta_i^{\mathrm{mis}}
	+ \sum_{k=1}^{2} \alpha_k\,\delta\theta_{i-k}^{\mathrm{mis}}.
\end{equation}
Here $\delta\theta_i^{\mathrm{mis}}$ is the angular deviation due to a mismatch at site~$i$, and the sum accounts for the influence of mismatches at the two neighboring positions, weighted by $\alpha_k$.

Applying Eyring transition-state theory and the Bell model, we obtain the following expressions for the rate constants of duplex closing and opening:
\begin{align}
	k^{i}_{\mathrm{on,1}} &= \kappa\,\frac{k_B T}{h}\,
	\exp\!\Big[-\Big(\Delta G^{\neq}_{1,i}
	- \Delta G^{i}_{\mathrm{sf1}} \nonumber\\
	&\qquad\qquad - \Delta W_i N_A
	+ P_2^{i}\,W^{\mathrm{on}}_{\mathrm{cas},i}\Big)/RT\Big],
	\label{k1on} \\
	k^{i}_{\mathrm{off,1}} &= \kappa\,\frac{k_B T}{h}\,
	\exp\!\Big[-\Big(\Delta G^{\neq}_{2,i}
	+ \Delta G^{i}_{\mathrm{sf1}} \nonumber\\
	&\qquad\qquad + \Delta W_i N_A
	- P_2^{i}\,W^{\mathrm{off}}_{\mathrm{cas},i}\Big)/RT\Big],
	\label{k1off}
\end{align}
where $\Delta G^{\neq}_{1,i}$ and $\Delta G^{\neq}_{2,i}$ are the activation free energies for base pair closing and opening, respectively. $\Delta G^{i}_{\mathrm{sf1}}$ is the base stacking contribution to the transition-state barrier. $W^{\mathrm{on}}_{\mathrm{cas},i}$ and $W^{\mathrm{off}}_{\mathrm{cas},i}$ are the works performed by Cas9 conformational changes during forward and reverse R-loop propagation, $N_A$ is Avogadro's number, $h$ is Planck's constant, and $\kappa$ is the transmission coefficient. We note that Eqs.~(\ref{k1on}) and~(\ref{k1off}) involve an implicit coarse graining. The actual work done by Cas9 during R-loop formation is assumed to be proportional to the binding probability $P_2^{i}$.

We now analyze the sgRNA-binding subprocess. As stated earlier, $P_2^{i}$ is defined under the condition that the target DNA strand at site~$i$ has already been exposed. The influence of the nontarget strand on binding is therefore neglected. The rate constants are
\begin{align}
	k^{i}_{\mathrm{on,2}} &= \kappa_{\mathrm{on}}\,\frac{k_B T}{h}\,
	\exp\!\left(-\frac{
		\Delta G^{\neq}_{3,i} - \Delta G^{i}_{\mathrm{sf2}}
	}{RT}\right),
	\label{k2on} \\
	k^{i}_{\mathrm{off,2}} &= \kappa_{\mathrm{off}}\,\frac{k_B T}{h}\,
	\exp\!\left(-\frac{
		\Delta G^{\neq}_{4,i} + \Delta G^{i}_{\mathrm{sf2}}
	}{RT}\right),
	\label{k2off}
\end{align}
where $\Delta G^{\neq}_{3,i}$ and $\Delta G^{\neq}_{4,i}$ are the activation free energies for binding and dissociation, and $\Delta G^{i}_{\mathrm{sf2}}$ is the base stacking barrier correction for this subprocess.

The equilibrium constant for the binding step is
\begin{equation}
	K_2^{i} = \frac{k^{i}_{\mathrm{on,2}}}{k^{i}_{\mathrm{off,2}}}
	= \exp\!\left(
	\frac{\Delta G^{\neq}_{4,i} - \Delta G^{\neq}_{3,i}
		+ 2\,\Delta G^{i}_{\mathrm{sf2}}}{RT}
	\right),
\end{equation}
and the corresponding equilibrium free-energy change satisfies $\Delta G^{\circ}_{i,\mathrm{rd}} = -RT\ln K_2^{i}$, yielding
\begin{equation}
	\Delta G^{\circ}_{i,\mathrm{rd}}
	= \Delta G^{\neq}_{3,i} - \Delta G^{\neq}_{4,i}
	- 2\,\Delta G^{i}_{\mathrm{sf2}}.
\end{equation}

We now consider the sequential kinetics of R-loop formation by the Cas9-sgRNA complex on the target DNA. Labeling the PAM-proximal end as $i=1$ and the distal end as $i=20$, each substep~$i$ involves two microscopic processes: (i)~unwinding and reclosing of the DNA duplex, and (ii)~base pairing between sgRNA and the complementary target strand. We denote the fundamental rate constants for duplex closing and opening at fragment~$i$ by $k^{i}_{\mathrm{on,1}}$ and $k^{i}_{\mathrm{off,1}}$, the sgRNA-binding probability by $P_2^{i}$, the effective Cas9 conformational works by $W^{\mathrm{on}}_{\mathrm{cas},i}$ and $W^{\mathrm{off}}_{\mathrm{cas},i}$, and the activation free energies for closing and opening by $\Delta G^{\neq}_{1,i}$ and $\Delta G^{\neq}_{2,i}$.

Eliminating the common prefactor from the forward and reverse transition-state expressions yields the thermodynamic form of the rate ratio at fragment~$i$:
\begin{widetext}
	\begin{equation}
		\ln\frac{k^{i}_{\mathrm{off,1}}}{k^{i}_{\mathrm{on,1}}}
		= \frac{
			\Delta G^{\neq}_{1,i} - \Delta G^{\neq}_{2,i}
			- 2\,\Delta G^{i}_{\mathrm{sf1}}
			- 2\,\Delta W_i N_A
			+ P_2^{i}\bigl(W^{\mathrm{on}}_{\mathrm{cas},i}
			+ W^{\mathrm{off}}_{\mathrm{cas},i}\bigr)
		}{RT},
		\label{eq:ratio_raw}
	\end{equation}
\end{widetext}
where $\Delta G^{i}_{\mathrm{sf1}}$ is the base stacking barrier correction, $\Delta W_i$ the torsional energy per substep, and $N_A$ Avogadro's number. For compactness, we introduce
\begin{equation}
	\Delta\Delta G^{\neq}_i \equiv
	\Delta G^{\neq}_{1,i} - \Delta G^{\neq}_{2,i},
	\qquad
	\mathcal{W}^{\mathrm{cas}}_i \equiv
	W^{\mathrm{on}}_{\mathrm{cas},i}
	+ W^{\mathrm{off}}_{\mathrm{cas},i},
\end{equation}
so that Eq.~(\ref{eq:ratio_raw}) becomes
\begin{equation}
	\ln\frac{k^{i}_{\mathrm{off,1}}}{k^{i}_{\mathrm{on,1}}}
	= \frac{
		\Delta\Delta G^{\neq}_i
		- 2\,\Delta G^{i}_{\mathrm{sf1}}
		- 2\,\Delta W_i N_A
		+ P_2^{i}\,\mathcal{W}^{\mathrm{cas}}_i
	}{RT}.
	\label{eq:ratio_compact}
\end{equation}
The factor $P_2^{i}$ reflects the cooperativity between Cas9 allosteric activation and DNA binding.

Defining the net driving work at a single site as
\begin{equation}
	W^{\mathrm{drive}}_i \equiv
	P_2^{i}\,\mathcal{W}^{\mathrm{cas}}_i
	- 2\,\Delta W_i N_A,
	\label{eq:Wdrive}
\end{equation}
Eq.~(\ref{eq:ratio_compact}) shows that the rate ratio is determined jointly by the energy difference and the external work. When the site is unbound ($P_2^{i}=0$), the conformational term vanishes and the ratio is governed solely by the activation energy difference, the stacking correction, and the torsional contribution.

Because DNA unwinding controls R-loop formation, we take the ratio of the forward to the reverse rate coefficient of the duplex at a single site as a measure of the local R-loop formation efficiency, $\eta_i$. Cleavage is not considered in this measure. Accounting for the fact that the transmission coefficients $\kappa$ differ between the forward and reverse directions, we obtain
\begin{equation}
	\eta_i \propto \exp\!\left[
	(RT)^{-1}\Big(
	\Delta\Delta G^{\neq}_i
	- 2\,\Delta G^{i}_{\mathrm{sf1}}
	+ W^{\mathrm{drive}}_i
	\Big)
	\right].
	\label{eq:eta}
\end{equation}

When the target DNA and the sgRNA are mismatched, three effects suppress the efficiency. First, a mismatch disrupts the hydrogen bond network and weakens stacking, which raises the local barrier $\Delta G^{i}_{\mathrm{sf1}}$. Second, the twist perturbation raises the torsional strain energy $\Delta W_i$, so that more work must be supplied against the junction. Third, the instability of the pair decreases both the bound probability $P_2^{i}$ and the allosteric work that Cas9 contributes. All three decrease $\eta_i$ and suppress R-loop formation.

The sgRNA-binding probability $P_2^{i}$ can be determined independently in the steady state. From the steady-state condition of Eq.~(\ref{p2}),
\begin{align}
	P_{2,\mathrm{ss}}^{i}
	&= \frac{k^{i}_{\mathrm{on,2}}}
	{k^{i}_{\mathrm{on,2}} + k^{i}_{\mathrm{off,2}}}
	\notag \\
	&= \left[1 + \exp\!\left(
	\frac{\Delta G^{\circ}_{i,\mathrm{rd}}}{RT}
	\right)\right]^{-1},
\end{align}
where $\Delta G^{\circ}_{i,\mathrm{rd}} = \Delta G^{\neq}_{3,i} - \Delta G^{\neq}_{4,i} - 2\,\Delta G^{i}_{\mathrm{sf2}}$ is the equilibrium free-energy change of the binding process. This probability directly modulates the driving work $\mathcal{W}^{\mathrm{cas}}_i$ supplied by Cas9 at fragment~$i$, establishing a feedback loop. Stronger binding provides more driving work, which promotes further DNA unwinding.

The model described above governs the kinetics of both DNA unwinding and RNA--DNA binding, with distinct rate constants for the forward and reverse processes. This asymmetry in the rate constants gives rise to entropy production, a key quantity characterizing the irreversibility of nonequilibrium systems. Quantifying the entropy production of R-loop formation can thus reveal the origin of the driving forces and the degree of irreversibility inherent in the process.

\section{Entropy production and irreversibility}
\label{sec:entropy}

To quantify the irreversibility of R-loop formation we use the entropy production $\hat{\sigma}$ along a trajectory that runs from the fully wound to the fully unwound state. For a forward path $\Gamma^+$ and its time reverse $\Gamma^-$,
\begin{equation}
	\hat{\sigma} = k_B \ln \left( \frac{P_+ [\Gamma_+]}{P_-[\Gamma_-]} \right)\label{eq:entropy_total}
\end{equation}
Because we consider trajectories that complete the R-loop, the path runs from $Q_0$ to $Q_{20}$. The dwell times cancel between the two path probabilities, so only the ratio of the rates along the sequence survives. For the entropy production, the contribution from system entropy change $k_B\ln[Q_0(0)/Q_{20}(T)]$ is relatively small compared with that of the medium entropy change, which is given by $k_B$ times the logarithm of the ratio of the forward and reverse transition rates. The derivation is given in Appendix~B and yields
\begin{equation}
	\hat{\sigma}= k_B \sum_{k=1}^{20} \ln\frac{k^{k}_{\mathrm{off},1}}{(1-\mu_{\mathrm{cleavage}}\delta_{k,20})\,k^{k}_{\mathrm{on},1}}, \label{eq:entropy_simple}
\end{equation}
so that
\begin{equation}
	\hat{\sigma} = k_B \sum_{k=1}^{19} \ln\eta_k
	+ k_B \ln\frac{\eta_{20}}{1-\mu_{\mathrm{cleavage}}}. \label{eq:S_eta}
\end{equation}
Using the expression for the rate ratio, Eq.~(\ref{eq:eta}),
\begin{equation}
	\ln\frac{k^{k}_{\mathrm{off},1}}{k^{k}_{\mathrm{on},1}}
	=(RT)^{-1}\Big[\Delta\Delta G^{\neq}_k - 2\Delta G^{k}_{\mathrm{sf1}}+W^{\mathrm{drive}}_k\Big],
\end{equation}
and substituting into Eq.~(\ref{eq:entropy_simple}), we obtain
\begin{widetext}
\begin{equation}
	\hat{\sigma} = \frac{k_B}{RT}\sum_{k=1}^{20}
	\bigl(\Delta\Delta G^{\neq}_k - 2\Delta G^{k}_{\mathrm{sf1}}+W^{\mathrm{drive}}_k\bigr)
	-k_B\ln(1-\mu_{\mathrm{cleavage}}). \label{eq:S_expanded}
\end{equation}
\end{widetext}
This expression separates four contributions of entropy production, the local activation free-energy difference, base stacking, the torsional energy, and the Cas9 allosteric work, the last two of which enter through $W^{\mathrm{drive}}_k$. A mismatch acts on these terms by raising $\Delta G^{k}_{\mathrm{sf1}}$ and $\Delta W_k$ and by lowering $P^k_2$, so that each mismatch makes the R-loop pathway less favourable. The entropy production framework thus provides a stochastic thermodynamic description for the editing specificity.

\section{Physically constrained activity prediction model}
\label{sec:physics_model}

The preceding sections provide a thermodynamic and kinetic framework for dCas9-sgRNA target binding. We now test it against experiment, with every energy scale either taken from measured structures or fixed by a calibration constant, and with no external thermodynamic database.

We use the data of Feng \textit{et al.}~\cite{Feng2021}, who coupled dCas9 binding to bacterial growth through a \textit{sacB} counterselection system and converted the growth-coupled retention into an enrichment per variant by high-throughput sequencing of the selected and control libraries. Replicate measurements of the r0 library reproduce with Pearson $r=0.91$~\cite{Feng2021}, so the measured activities are reproducible enough to serve as a quantitative target. Two libraries are analyzed, r0 and r349, supplied as the original workbooks \texttt{Binding\_activity\_r0.xlsx} and \texttt{Binding\_activity\_r349.xlsx}. The r0 library contains the wild type plus $1704$ mutants, of which $60$ carry a single mismatch and $1644$ a double mismatch. Its wild-type sgRNA is \texttt{UACACUUGAACUACCGCGAG}, and its DNA target strand is \texttt{ATGTGAACTTGATGGCGCTC}. The r349 library contains the wild type plus $1718$ mutants, of which $60$ carry a single mismatch and $1658$ a double mismatch. The activity spans $-8.494$ to $5.420$ for r0, with mean $-2.758$ and standard deviation $1.952$, and $-9.764$ to $4.380$ for r349, with mean $-3.728$ and standard deviation $2.182$. Feng \textit{et al.} define the activity as the $\log_2$ enrichment relative to the wild type,
\begin{align}
	\mathrm{Fitness}_i&=\log_2\frac{\mathrm{Read\ count}_{i,\mathrm{selective}}}{\mathrm{Read\ count}_{i,\mathrm{control}}},\nonumber\\
	\mathrm{Binding\ activity}_i&=\mathrm{Fitness}_i-\mathrm{Fitness}_{\mathrm{WT}},
	\label{eq:fitness}
\end{align}
and the \texttt{score} column of the workbooks stores this quantity directly, with the wild type entering as a baseline of zero, so that the two libraries comprise $1705$ (r0) and $1719$ (r349) variants. Throughout this section positions are numbered $p=1,\ldots,20$ from the PAM-distal to the PAM-proximal end, so that $p=20$ lies next to the PAM. This is the numbering of the dataset, and it runs opposite to the index $i$ of Secs.~\ref{sec:model}--\ref{sec:entropy}. Mismatch types follow the notation of Feng \textit{et al.}, dXrY, where X is the target DNA base and Y the sgRNA base.

The enrichment is a product over positions. Because dCas9 retains variant $i$ with the probability that a complete R-loop forms, and because R-loop propagation is directional~\cite{Sternberg2014,Szczelkun2014,Gong2018}, the retention factorizes over the twenty positions,
\begin{equation}
	\frac{N_{i,\mathrm{selective}}}{N_{i,\mathrm{control}}}\propto\prod_{p=1}^{20}\eta_p,
	\qquad
	\mathrm{score}_i\propto\sum_{p=1}^{20}\log_2\eta_p,
	\label{eq:score_decomposition}
\end{equation}
where $\eta_p$ is the local probability that position $p$ completes its RNA--DNA hybrid. Up to an additive wild-type constant, the measured activity is therefore a sum of per position logarithmic formation efficiencies, which allows it to be replaced by a free-energy sum.

Unwinding each base pair also changes the twist of the surrounding duplex by the helical twist of B-DNA, $\approx34.3^{\circ}$ per base pair~\cite{Ivanov2020,Watson1953}, so that accumulated twist stress grows with the number of positions that complete their hybrid. We therefore let the free-energy cost of a mismatch be carried by the local twist deviation rather than by the base pair hydrogen bonds alone,
\begin{equation}
	\mathrm{score}\propto\Delta G_{\mathrm{total}}=\sum_{p=1}^{20}\Delta G_p,
	\label{eq:score_free_energy}
\end{equation}
which we treat as the central hypothesis of the model. This is where the model departs from Feng \textit{et al.}~\cite{Feng2021}, who decompose the same activity into nearest neighbor base pairing thermodynamics, with parameters from the DNA:RNA literature~\cite{Sugimoto1995,Watkins2011,Lesnik1995}, plus a global Cas9 compensation of $-3000$~J/mol/base. That thermodynamic model can be evaluated only for the $\approx9.2\%$ of mutants whose mismatch types have tabulated parameters, so Feng \textit{et al.} additionally trained a convolutional neural network covering all mutants. The present construction keeps the interpretability of the free-energy approach while covering every mutant with a small set of physical parameters.

The activity of variant $i$ is
\begin{equation}
	S_i=b_{\mathrm{rd}}+c\sum_{p=1}^{20}\bigl[w_{\mathrm{cas},p}\,g_p-\Delta W_p\bigr],
	\label{eq:model}
\end{equation}
where $b_{\mathrm{rd}}$ is an intercept, $c$ a global scale, $w_{\mathrm{cas},p}=\gamma_p W_{\mathrm{Cas}}^{\mathrm{init}}$ with $W_{\mathrm{Cas}}^{\mathrm{init}}=0.055$, $g_p$ the pairing weight, and $\Delta W_p$ the torsional cost. The pairing weight is unity at Watson--Crick-matched positions and equals $[1+\exp(-w_{\mathrm{rd}})]^{-1}$ at a mismatch, with $w_{\mathrm{rd}}$ a weight that depends on the mismatch type. The torsional cost is
\begin{align}
	\Delta W_p&=\frac{C\,(\Delta\theta_p)^2}{\Delta x},\nonumber\\
	\Delta\theta_p&=\Delta\theta_0+\delta\theta_p^{\mathrm{mis}}+\alpha_1\,\delta\theta_{p-1}^{\mathrm{mis}}+\alpha_2\,\delta\theta_{p-2}^{\mathrm{mis}},
	\label{eq:twist_energy}
\end{align}
with $C=1.4169\times10^{-6}$, $\Delta x=1$, and $\Delta\theta_0=1.56^{\circ}=27.20$~mrad. Energies are expressed in $k_BT$, angles in mrad, and lengths in base pairs. The trainable parameters are the $12$ values of $w_{\mathrm{rd}}$, the $12$ values of $\delta\theta$, the $20$ values of $\gamma_p$, and $b_{\mathrm{rd}},c,\alpha_1,\alpha_2$.

The energy of Eq.~(\ref{eq:twist_energy}) is the elastic rod twist energy, and $C$ is a torsional modulus in units of (mrad)$^{-2}$ whose value is fixed in the model rather than fitted to the activity data. We therefore treat $C$ as a calibration of the energy scale within this coarse-grained model, not as an independently measured DNA constant. The offset $\Delta\theta_0=1.56^{\circ}$ is the twist difference between the A-form-like RNA--DNA hybrid, with $11.0$ bp per turn ($32.7^{\circ}$ per base pair), and the flanking B-DNA, with $10.5$ bp per turn ($34.3^{\circ}$ per base pair)~\cite{Sugimoto1995}. The coefficients $\alpha_1,\alpha_2$ describe how a twist deformation propagates to neighboring positions instead of remaining local~\cite{Yakovchuk2006}. The scale $W_{\mathrm{Cas}}^{\mathrm{init}}=0.055$ fixes the size of the allosteric term so that the fitted work per position, $c\,\gamma_p W_{\mathrm{Cas}}^{\mathrm{init}}$, stays in the range of the $-1.2\,k_BT$ per base compensation of Feng \textit{et al.}~\cite{Feng2021} and of the $2$--$6\,k_BT$/bp measured for R-loop formation in single-molecule experiments~\cite{Ivanov2020,Szczelkun2014}. The training protocol, the split of the data, the initialization, and the constraints imposed on the parameters are described in Sec.~S1 of the Supplemental Material.

\begin{figure}[!htbp]
\centering
\includegraphics[width=0.85\linewidth]{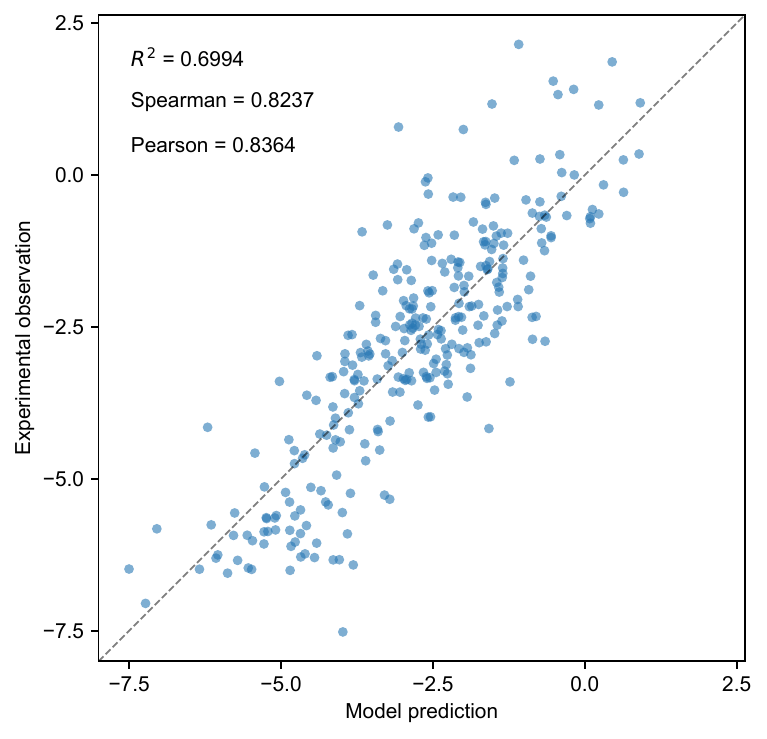}
\caption{Predicted against measured binding activity on the r0 test set ($n=302$). The dashed line is $y=x$.}
\label{fig:prediction_r0}
\end{figure}

\begin{figure}[!htbp]
\centering
\includegraphics[width=0.85\linewidth]{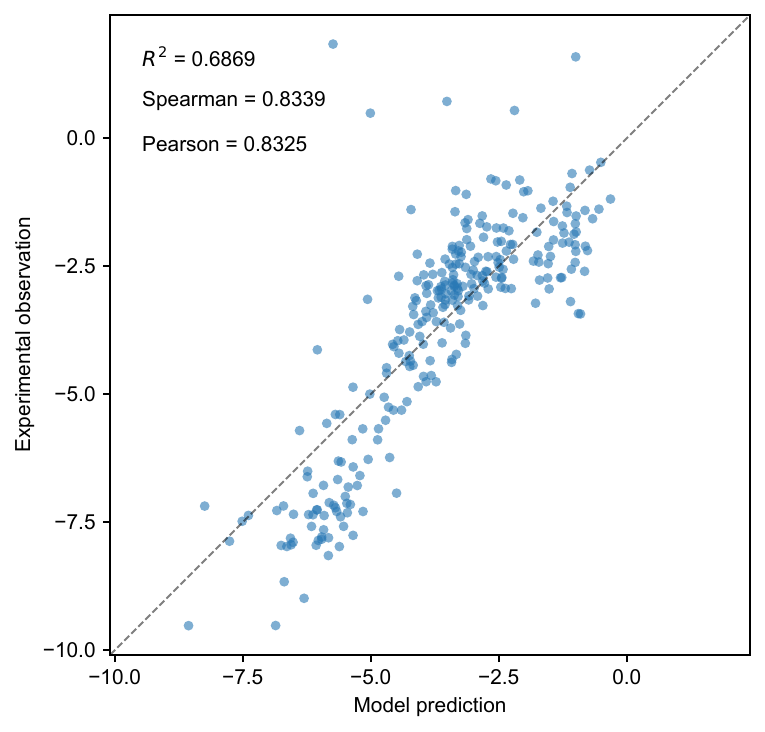}
\caption{Predicted against measured binding activity on the r349 test set ($n=275$). The dashed line is $y=x$.}
\label{fig:prediction_r349}
\end{figure}

On the held-out test sets the model reaches $R^2=0.68\pm0.06$ and $\mathrm{Spearman}=0.84\pm0.03$ for r0 and $R^2=0.64\pm0.05$ and $\mathrm{Spearman}=0.82\pm0.03$ for r349, averaged over seven independent random partitions (seeds $0$ to $6$). For the representative partition used below (seed $0$) r0 reaches $R^2=0.6994$, $\mathrm{RMSE}=1.047$, $\mathrm{Spearman}=0.8237$, $\mathrm{Pearson}=0.8364$ and r349 reaches $R^2=0.6869$, $\mathrm{RMSE}=1.236$, $\mathrm{Spearman}=0.8339$, $\mathrm{Pearson}=0.8325$ (Figs.~\ref{fig:prediction_r0} and~\ref{fig:prediction_r349}). Both rank correlations exceed the Spearman correlation of $0.64$ reported by Feng \textit{et al.} for their thermodynamic model~\cite{Feng2021}. Part of that gap is physical, in that the present model resolves position and mismatch type into separate parameter sets, and part is methodological, since Feng \textit{et al.} use a random $80/20$ split whereas here a fixed coverage split is used, so the comparison is indicative rather than a controlled benchmark. The variation of the test metrics over the seven partitions, and the effect of excluding the most active variants from the fit, are reported in Sec.~S2 and Fig.~S1 of the Supplemental Material.

The residuals are structured rather than uniform, and the structure is informative. Across the seven partitions the test residuals of both libraries carry a heavy left tail, with excess kurtosis between $0.9$ and $17.6$ for r0 and between $1.7$ and $15.0$ for r349, and the flagged variants are almost exclusively double mismatches of high measured activity that the model underpredicts (Sec.~S3 and Figs.~S2--S9 of the Supplemental Material). Training the same model on the subset of nonpositive activity, which holds $94\%$ of the r0 and $99\%$ of the r349 variants, restores Gaussian-like residuals, whereas the positive-activity, double mismatch tail remains underpredictive (Figs.~S10--S13 of the Supplemental Material). The tail therefore marks the regime in which the present free-energy mapping is inadequate, for two physical reasons. First, two nearby mismatches are not independent, and a form that multiplies per position efficiencies, equivalently that adds per position free energies, cannot represent the synergy by which such a variant keeps its activity. The coefficients $\alpha_k$ absorb only part of that coupling. Second, part of the cost of a double mismatch can be relaxed locally. The Arg71--guanine contact in Cas9~\cite{Anders2014}, together with the high stability of G-containing noncanonical pairs~\cite{Allawi1997}, provides a structural route by which mismatches involving G cost less than their tabulated thermodynamic parameters suggest, in the same direction as the small twist deviations fitted for the dDrG class below.

\begin{figure}[!htbp]
	\centering
	\includegraphics[width=0.92\linewidth]{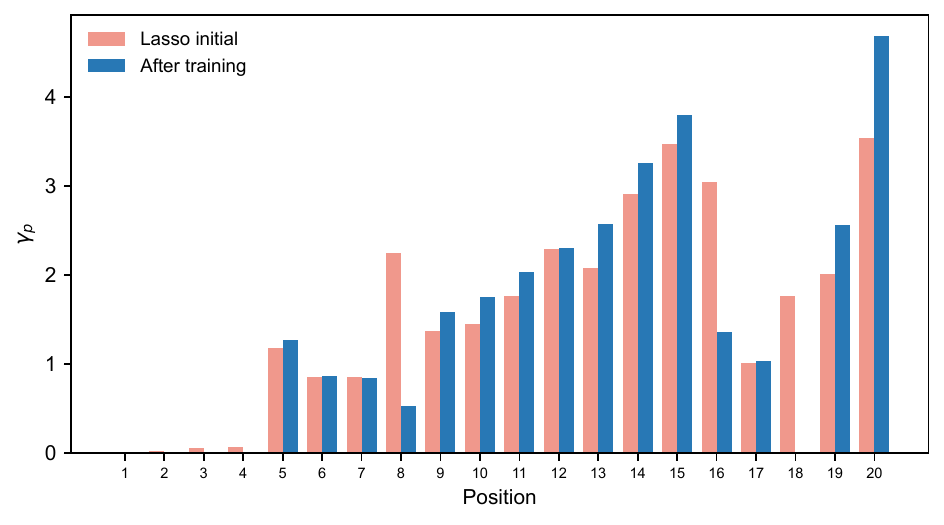}
	\caption{Initial and fitted position coefficients $\gamma_p$ (r0).}
	\label{fig:position}
\end{figure}

The fitted parameters reproduce the experimental phenomenology and expose library-specific structure. For the representative partition the coefficients $\gamma_p$ vanish at positions $1$ to $4$, rise across positions $5$ to $14$, and are largest near the PAM-proximal end, with $\gamma_{15}=3.79$, $\gamma_{19}=2.56$, and $\gamma_{20}=4.69$ for r0 (Fig.~\ref{fig:position}). Since position $20$ is the PAM-proximal end in this numbering, the largest penalties lie in the seed and PAM regions, which is consistent with a directional R-loop in which a mismatch encountered early blocks propagation and costs more than one encountered late, and it agrees with the seed sensitivity and distal tolerance reported elsewhere~\cite{Feng2021,Sternberg2014,Hsu2013,Fu2013,Pattanayak2013}.

\begin{figure}[!htbp]
	\centering
	\includegraphics[width=0.92\linewidth]{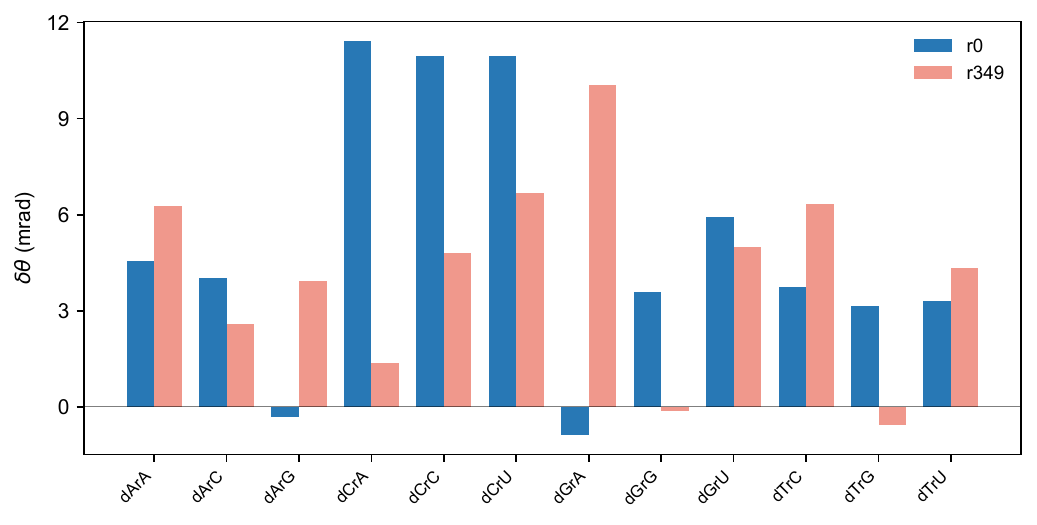}
	\caption{Fitted twist deviations $\delta\theta$ for each mismatch type, r0 and r349.}
	\label{fig:type_delta}
\end{figure}

For the representative partition the twist deviations of the dDrG class stay small in most cases, with dArG$=-0.32$, dGrG$=3.58$, and dTrG$=3.14$~mrad for r0 and dArG$=3.92$, dGrG$=-0.13$, and dTrG$=-0.56$~mrad for r349, which parallels the moderate activity loss caused by dDrG mismatches~\cite{Feng2021} (Fig.~\ref{fig:type_delta}). The largest deviations are instead library-specific, with dCrA$=11.42$, dCrC$=10.96$, and dCrU$=10.95$~mrad dominating in r0 whereas dGrA dominates in r349 at $10.04$~mrad and the dCr values are only moderate at $1.38$, $4.81$, and $6.69$~mrad. The shared smallness of dDrG is thus a robust signature, whereas the ranking of the remaining types does not transfer between libraries. Within the model, a small or negative $\delta\theta$ means that these mismatches barely perturb the hybrid twist and carry a small deformation cost, and a negative sign indicates a deviation opposite to $\Delta\theta_0$, that is, a partial compensation of the hybrid duplex twist mismatch, which would be consistent with the Arg71--guanine contact~\cite{Anders2014}. Because for the dDrG class these magnitudes are of order $1$~mrad, close to the resolution of the fit, we regard the sign as qualitative.

Finally, the twist propagation is strongly asymmetric, with $\alpha_1=22.48$ and $\alpha_2=5.7\times10^{-4}$ for r0, together with $c=19.47$ and $b_{\mathrm{rd}}=-31.35$ for r0 and $c=17.70$ and $b_{\mathrm{rd}}=-33.65$ for r349. A large $\alpha_1$ with a vanishing $\alpha_2$ means that the deformation propagates essentially one position and not two, which offers a mechanism for the seed synergy observed by Feng \textit{et al.}~\cite{Feng2021}, while the negligible $\alpha_2$ indicates that the coupling is short-ranged within the accuracy of the fit.

Taken together, these results show that a physically constrained torsional model, without any external thermodynamic database, reproduces the two libraries to $R^2=0.68\pm0.06$ and $0.64\pm0.05$ over seven independent partitions, recovers the seed and PAM gradient together with the moderate cost of dDrG mismatches, and isolates the high-activity tail as the regime in which the current free-energy mapping is inadequate.

\section{Conclusions}
\label{sec:conclusion}
	
This work establishes a nonequilibrium analytic framework for studying R-loop formation in the CRISPR-Cas9 system, based on a master equation combined with thermodynamic analysis, and quantifies the irreversibility of the process through its entropy production. On this basis we construct a prediction model that is trained on experimental data, uses no external thermodynamic database, and reproduces the binding activity of both libraries with rank correlations of $0.84$ and $0.82$. The comparison provides a first validation of the framework and makes the model parameters interpretable in physical terms.
	
We would like to note that the position dependence of mismatch tolerance emerges from the sequential structure of R-loop propagation, in which each step depends on the completion of the preceding one, and the largest penalties fall on the seed and PAM regions. The fitted twist deviations for dDrG mismatches are small in both libraries, consistent with the moderate activity loss that this class shows experimentally. The residuals are largest where double mismatches in close sequence proximity keep their binding activity, so the per position product form does not fully capture the interaction of nearby mismatches. Part of the residual may also reflect the finite sample size, which limits how precisely model parameters can be determined.
	
Two directions follow. First, this work assumes that each position contributes independently to the binding probability, and the twist propagation coefficients $\alpha_1$ and $\alpha_2$ correct this assumption only for twist deformation propagated to the two nearest neighbors. Double mismatches separated by more than two positions are not captured by this correction, which is the possible source of the residual heavy tail in the r349 library. A more general coupling structure, such as a position dependent interaction range or a many-body term, would address this. Second, the master equation formulation is transferable to other CRISPR effector complexes. Cas12a, for example, also forms an R-loop through sequential unwinding from the PAM-proximal end, but its nuclease activation mechanism and PAM requirements differ from those of Cas9. Adapting the framework to Cas12a would allow direct comparison of the fitted torsional, stacking, and allosteric parameters between the two systems, revealing whether the differences in mismatch tolerance between Cas9 and Cas12a arise from differences in these specific physical quantities.

\clearpage
\appendix
\section{Derivation of the transition rates}
\label{app:A}

Here we show that the master equation, Eq.~(\ref{eq:master}), reduces to Eq.~(\ref{dotP}) once the cumulative probabilities $P_1^{i}$ are introduced, and we read off the transition rates.

From Eqs.~(\ref{eq:P1Q}) and~(\ref{eq:master}),
\begin{equation}
	\begin{aligned}
		\dot{P}_1^{i}&= \sum_{m=i}^{20} \dot{Q}_m \\
		&= \sum_{m=i}^{19}\Big[
		\lambda_{m,m-1}\,Q_{m-1}
		- (\lambda_{m-1,m} + \lambda_{m+1,m})\,Q_m \\
		&\qquad + \lambda_{m,m+1}\,Q_{m+1}
		\Big] \\
		&\qquad + \lambda_{20,19}\,Q_{19}
		- \lambda_{19,20}\,Q_{20}.
	\end{aligned}
\end{equation}
Expanding the sum,
\begin{equation}
	\begin{aligned}
		\dot{P}_1^{i}
		&= \lambda_{i,i-1}\,Q_{i-1}
		- (\lambda_{i-1,i} + \lambda_{i+1,i})\,Q_i
		+ \lambda_{i,i+1}\,Q_{i+1} \\
		&\quad + \lambda_{i+1,i}\,Q_i
		- (\lambda_{i,i+1} + \lambda_{i+2,i+1})\,Q_{i+1} \\
		&\quad + \lambda_{i+1,i+2}\,Q_{i+2} \\
		&\quad + \cdots \\
		&\quad + \lambda_{20,19}\,Q_{19}
		- \lambda_{19,20}\,Q_{20}.
	\end{aligned}
\end{equation}
Pairwise cancellation of adjacent terms yields
\begin{equation}
	\begin{aligned}
		\dot{P}_1^{i} &= 0, \quad (i = 0), \\
		\dot{P}_1^{i} &= \lambda_{i,i-1}\,Q_{i-1}
		- \lambda_{i-1,i}\,Q_i, \quad (1 \leq i \leq 20).
	\end{aligned}
	\label{eq:mid}
\end{equation}

Substituting Eq.~(\ref{eq:QK119}) into Eq.~(\ref{eq:mid}) and dividing both sides by $P_1^{i-1}$,
\begin{equation}
	\frac{\dot{P}_1^{i}}{P_1^{i-1}}
	= \lambda_{i,i-1}\,
	\frac{P_1^{i-1} - P_1^{i}}{P_1^{i-1}}
	- \lambda_{i-1,i}\,
	\frac{P_1^{i} - P_1^{i+1}}{P_1^{i-1}}.
\end{equation}
Inserting $q_i = P_1^{i}/P_1^{i-1}$ and $q_{i+1} = P_1^{i+1}/P_1^{i}$, we obtain
\begin{equation}
	\frac{\dot{P}_1^{i}}{P_1^{i-1}}
	= \lambda_{i,i-1}\,(1 - q_i)
	- \lambda_{i-1,i}\,q_i\,(1 - q_{i+1}).
\end{equation}
Comparing coefficients with Eq.~(\ref{dotP}) yields the consistency conditions, from which we obtain
\begin{equation}
	\lambda_{i,i-1}=k^{i}_{\mathrm{off,1}},
	\qquad
	\lambda_{i-1,i}=(1-\mu_{\mathrm{cleavage}}\delta_{i,20})k^{i}_{\mathrm{on,1}}.
\end{equation}

\section{Entropy production of the sequential R-loop process}\label{app:B}

Within the framework of stochastic thermodynamics, the irreversibility of R-loop formation is quantified by the total entropy production along a trajectory of the sequential process $0 \to 1 \to \cdots \to 20$,
\begin{equation}
	\hat{\sigma} = k_B \ln \frac{P_{+}[\Gamma^{+}]}{P_{-}[\Gamma^{-}]}, \label{eq:entropy_def}
\end{equation}
where $P_{+}$ and $P_{-}$ are the probability densities of the forward trajectory $\Gamma^{+}$ and of its time reversal $\Gamma^{-}$, respectively.

Consider a forward trajectory that starts in state~$0$, visits the states in order, and is terminated in state~$20$ by cleavage. Let $\tau_i$ ($i = 0, 1, \ldots, 19$) denote the waiting time spent in state~$i$ before the jump $i \to i{+}1$. The dwell in state~$i$ contributes the exponential factor $e^{-r_i \tau_i}$, with $r_i$ the total exit rate from state~$i$, and the jump $i \to i{+}1$ contributes the rate $\lambda_{i+1,i}$. Hence
\begin{equation}P_{+}(\Gamma^{+})
	= \left(\prod_{i=1}^{20} \lambda_{i,i-1}\right) \exp\!\left(-\sum_{i=0}^{19} r_i \tau_i\right),\label{eq:Pplus}
\end{equation}
with the exit rates
\begin{equation}
	r_0 = \lambda_{1,0}, \qquad
	r_i = \lambda_{i-1,i} + \lambda_{i+1,i} \quad (i = 1, 2, \ldots, 19).\label{eq:exit_rates}
\end{equation}
Each state~$i \geq 1$ has two escape channels, the forward jump $i \to i{+}1$ and the backward jump $i \to i{-}1$, both of which are counted in Eq.~\eqref{eq:exit_rates}.

On the time-reversed trajectory the jumps occur with the reversed rates $\lambda_{i-1,i}$, while time reversal preserves the sequence of dwell times and the exit rates. The waiting time factors in the reversed probability density are therefore identical to those in Eq.~\eqref{eq:Pplus}. Taking the reversed trajectory to start from state~$20$ with occupation $Q_{20}$,
\begin{equation}P_{-}(\Gamma^{-})= Q_{20}\left(\prod_{i=1}^{20} \lambda_{i-1,i}\right)\exp\!\left(-\sum_{i=0}^{19} r_i \tau_i\right). \label{eq:Pminus}
\end{equation}

Inserting Eqs.~\eqref{eq:Pplus} and~\eqref{eq:Pminus} into Eq.~\eqref{eq:entropy_def}, the waiting time factors cancel identically, and the entropy production separates into a system part and a medium part,
\begin{equation}
	\hat{\sigma} = k_B \ln \frac{Q_0(0)}{Q_{20}(T)}+{k_B \sum_{i=1}^{20}\ln \frac{\lambda_{i,i-1}}{\lambda_{i-1,i}}}.\label{eq:S_split}
\end{equation}
Here $Q_0(0)$ is the occupation of state~$0$ at the initial time and $Q_{20}(T)$ that of state~$20$ at the terminal time~$T$.

For a Cas9 complex that is catalytically competent the cleavage efficiency is high, so a completed R-loop is cleared by cleavage rather than by dissociation, and for the mismatched targets considered here the R-loop, once formed, is strongly driven forward. After a long time $T$ from $Q_0(0)$ the system therefore sits at $Q_{20}(T)$, and the system term $k_B\ln[Q_0(0)/Q_{20}(T)]$ is negligible compared with the medium term.

Substituting the rate identifications of Eqs.~(\ref{eq:lambda_full}) and~(\ref{eq:mu_full}), valid for $1 \leq i \leq 20$, one has
\begin{equation}
	\ln \frac{\lambda_{i,i-1}}{\lambda_{i-1,i}}= \ln \frac{k^{i}_{\mathrm{off},1}}{\left(1 - \mu_{\mathrm{cleavage}}\,\delta_{i,20}\right) k^{i}_{\mathrm{on},1}},\label{eq:ln_lambda_mu_expand}
\end{equation}
and summing over~$i$ gives
\begin{equation}
	\hat{\sigma}= k_B \sum_{i=1}^{20}\ln \frac{k^{i}_{\mathrm{off},1}}{\left(1 - \mu_{\mathrm{cleavage}}\,\delta_{i,20}\right) k^{i}_{\mathrm{on},1}}.\label{eq:S_med_expanded}
\end{equation}
Since the Kronecker symbol singles out $i = 20$, the sum splits into the nineteen reversible transitions and the terminal transition, at which cleavage occurs,
\begin{equation}
	\hat{\sigma}= k_B\sum_{i=1}^{20} \ln \frac{k^{i}_{\mathrm{off},1}}{k^{i}_{\mathrm{on},1}}-k_B\ln(1- \mu_{\mathrm{cleavage}}).\label{eq:S_med_split}
\end{equation}

\begin{acknowledgments}
This work was supported by the National Natural Science Foundation of China (NSFC) under Grant No.~12575045.
\end{acknowledgments}

\section*{Data availability}
The data that support the findings of this article are not publicly available. The data are available from the authors upon reasonable request.


\bibliography{refs}

@article{Feng2021,
  title   = {Guide-target mismatch effects on {dCas9}--{sgRNA} binding activity in living bacterial cells},
  author  = {Feng, Huibao and Guo, Jiahui and Wang, Tianmin and Zhang, Chong and Xing, Xin-Hui},
  journal = {Nucleic Acids Research},
  volume  = {49}, number = {3}, pages = {1263--1277}, year = {2021}, doi = {10.1093/nar/gkaa1295}
}

@article{Ivanov2020,
  title   = {{Cas9} interrogates {DNA} in discrete steps modulated by mismatches and supercoiling},
  author  = {Ivanov, Ivan E. and Wright, Addison V. and Cofsky, Joshua C. and Palacio Aris, Kevin D. and Doudna, Jennifer A. and Bryant, Zev},
  journal = {Proceedings of the National Academy of Sciences},
  volume  = {117}, number = {11}, pages = {5853--5860}, year = {2020}, doi = {10.1073/pnas.1913445117}
}

@article{Allawi1997,
  title   = {Thermodynamics and {NMR} of internal {G}$\cdot${T} mismatches in {DNA}},
  author  = {Allawi, Hatim T. and SantaLucia, John},
  journal = {Biochemistry},
  volume  = {36}, number = {34}, pages = {10581--10594}, year = {1997}, doi = {10.1021/bi962590c}
}

@article{Sternberg2014,
  title   = {{DNA} interrogation by the {CRISPR} {RNA}-guided endonuclease {Cas9}},
  author  = {Sternberg, Samuel H. and Redding, Sy and Jinek, Martin and Greene, Eric C. and Doudna, Jennifer A.},
  journal = {Nature},
  volume  = {507}, number = {7490}, pages = {62--67}, year = {2014}, doi = {10.1038/nature13011}
}

@article{Szczelkun2014,
  title   = {Direct observation of {R}-loop formation by single {RNA}-guided {Cas9} and {Cascade} effector complexes},
  author  = {Szczelkun, Mark D. and Tikhomirova, Maria S. and Sinkunas, Tomas and Gasiunas, Giedrius and Karvelis, Tautvydas and Pschera, Patrizia and Siksnys, Virginijus and Seidel, Ralf},
  journal = {Proceedings of the National Academy of Sciences},
  volume  = {111}, number = {27}, pages = {9798--9803}, year = {2014}, doi = {10.1073/pnas.1402597111}
}

@article{Gong2018,
  title   = {{DNA} unwinding is the primary determinant of {CRISPR}-{Cas9} activity},
  author  = {Gong, Shanzhong and Yu, Helen Hong and Johnson, Kenneth A. and Taylor, David W.},
  journal = {Cell Reports},
  volume  = {22}, number = {2}, pages = {359--371}, year = {2018}, doi = {10.1016/j.celrep.2017.12.041}
}

@article{Anders2014,
  title   = {Structural basis of {PAM}-dependent target {DNA} recognition by the {Cas9} endonuclease},
  author  = {Anders, Carolin and Niewoehner, Ole and Duerst, Alessia and Jinek, Martin},
  journal = {Nature},
  volume  = {513}, number = {7519}, pages = {569--573}, year = {2014}, doi = {10.1038/nature13579}
}

@article{Hsu2013,
  title   = {{DNA} targeting specificity of {RNA}-guided {Cas9} nucleases},
  author  = {Hsu, Patrick D. and Scott, David A. and Weinstein, Joshua A. and Ran, F. Ann and Konermann, Silvana and Agarwala, Vineeta and Li, Yinqing and Fine, Eli J. and Wu, Xuebing and Shalem, Ophir and Cradick, Thomas J. and Marraffini, Luciano A. and Bao, Gang and Zhang, Feng},
  journal = {Nature Biotechnology},
  volume  = {31}, number = {9}, pages = {827--832}, year = {2013}, doi = {10.1038/nbt.2647}
}

@article{Fu2013,
  title   = {High-frequency off-target mutagenesis induced by {CRISPR}-{Cas} nucleases in human cells},
  author  = {Fu, Yanfang and Foden, Jennifer A. and Khayter, Cyd and Maeder, Morgan L. and Reyon, Deepak and Joung, J. Keith and Sander, Jeffry D.},
  journal = {Nature Biotechnology},
  volume  = {31}, number = {9}, pages = {822--826}, year = {2013}, doi = {10.1038/nbt.2623}
}

@article{Pattanayak2013,
  title   = {High-throughput profiling of off-target {DNA} cleavage reveals {RNA}-programmed {Cas9} nuclease specificity},
  author  = {Pattanayak, Vikram and Lin, Steven and Guilinger, John P. and Ma, Enbo and Doudna, Jennifer A. and Liu, David R.},
  journal = {Nature Biotechnology},
  volume  = {31}, number = {9}, pages = {839--843}, year = {2013}, doi = {10.1038/nbt.2673}
}

@article{Sugimoto1995,
  title   = {Thermodynamic parameters to predict stability of {RNA}/{DNA} hybrid duplexes},
  author  = {Sugimoto, Naoki and Nakano, Shu-ichi and Katoh, Misa and Matsumura, Akiko and Nakamuta, Hiroyuki and Ohmichi, Tatsuo and Yoneyama, Mari and Sasaki, Muneo},
  journal = {Biochemistry},
  volume  = {34}, number = {35}, pages = {11211--11216}, year = {1995}, doi = {10.1021/bi00035a029}
}

@article{Lesnik1995,
  title   = {Relative thermodynamic stability of {DNA}, {RNA}, and {DNA}:{RNA} hybrid duplexes},
  author  = {Lesnik, Elena A. and Freier, Susan M.},
  journal = {Biochemistry},
  volume  = {34}, number = {34}, pages = {10807--10815}, year = {1995}, doi = {10.1021/bi00034a013}
}

@article{Watkins2011,
  title   = {Thermodynamic contributions of single internal {rA}$\cdot${dA}, {rC}$\cdot${dC}, {rG}$\cdot${dG} and {rU}$\cdot${dT} mismatches in {RNA}/{DNA} duplexes},
  author  = {Watkins, Norman E. and Kennelly, William J. and Tsay, Mu-Jing and Tuin, Astrid and Swenson, Lars and Lee, Hae-Ryong and Morosyuk, Sergiy and Weng, Kuei-Ming and SantaLucia, John},
  journal = {Nucleic Acids Research},
  volume  = {39}, number = {5}, pages = {1894--1902}, year = {2011}, doi = {10.1093/nar/gkq905}
}

@article{Jinek2012,
  title   = {A programmable dual-{RNA}-guided {DNA} endonuclease in adaptive bacterial immunity},
  author  = {Jinek, Martin and Chylinski, Krzysztof and Fonfara, Ines and Hauer, Michael and Doudna, Jennifer A. and Charpentier, Emmanuelle},
  journal = {Science},
  volume  = {337}, number = {6096}, pages = {816--821}, year = {2012}, doi = {10.1126/science.1225829}
}

@article{Nishimasu2014,
  title   = {Crystal structure of {Cas9} in complex with guide {RNA} and target {DNA}},
  author  = {Nishimasu, Hiroshi and Ran, F. Ann and Hsu, Patrick D. and Konermann, Silvana and Shehata, Shehata I. and Dohmae, Naoshi and Ishitani, Ryuichiro and Zhang, Feng and Nureki, Osamu},
  journal = {Cell},
  volume  = {156}, number = {5}, pages = {935--949}, year = {2014}, doi = {10.1016/j.cell.2014.02.001}
}

@article{Semenova2011,
  title   = {Interference by clustered regularly interspaced short palindromic repeat ({CRISPR}) {RNA} is governed by a seed sequence},
  author  = {Semenova, Ekaterina and Jore, Matthijs M. and Datsenko, Konstantin A. and Semenova, Anna and Westra, Edze R. and Wanner, Barry and van der Oost, John and Brouns, Stan J. J. and Severinov, Konstantin},
  journal = {Proceedings of the National Academy of Sciences},
  volume  = {108}, number = {25}, pages = {10098--10103}, year = {2011}, doi = {10.1073/pnas.1103090108}
}

@article{Boyle2017,
  title   = {High-throughput biochemical profiling reveals sequence determinants of {dCas9} off-target binding and unbinding},
  author  = {Boyle, Ethan A. and Andreasson, Jesper O. L. and Chircus, Lewis M. and Sternberg, Samuel H. and Wu, Michael J. and Guegler, Christian K. and Doudna, Jennifer A. and Greenleaf, William J.},
  journal = {Proceedings of the National Academy of Sciences},
  volume  = {114}, number = {21}, pages = {5461--5466}, year = {2017}, doi = {10.1073/pnas.1700557114}
}

@article{Doench2016,
  title   = {Optimized {sgRNA} design to maximize activity and minimize off-target effects of {CRISPR}-{Cas9}},
  author  = {Doench, John G. and Fusi, Nicolo and Sullender, Meredith and Hegde, Mudra and Vaimberg, Emma W. and Donovan, Katherine F. and Smith, Ian and Tothova, Zuzana and Wilen, Craig and Orchard, Robert and Virgin, Herbert W. and Liang, Xuebing and Bhatt, Deepak K. and Root, David E. and Hahn, William C.},
  journal = {Nature Biotechnology},
  volume  = {34}, number = {2}, pages = {184--191}, year = {2016}, doi = {10.1038/nbt.3437}
}

@article{MorenoMateos2015,
  title   = {{CRISPRscan}: designing highly efficient {sgRNAs} for {CRISPR}-{Cas9} targeting in vivo},
  author  = {Moreno-Mateos, Miguel A. and Vejnar, Charles E. and Beaudoin, Jean-Denis and Fernandez, Jaime P. and Mis, Eve K. and Khokha, Mustafa K. and Giraldez, Antonio J.},
  journal = {Nature Methods},
  volume  = {12}, number = {10}, pages = {982--988}, year = {2015}, doi = {10.1038/nmeth.3543}
}

@article{Doudna2014,
  title   = {The new biology of {Cas9} and its siblings},
  author  = {Doudna, Jennifer A. and Charpentier, Emmanuelle},
  journal = {Science},
  volume  = {346}, number = {6213}, pages = {1258096}, year = {2014}, doi = {10.1126/science.1258096}
}

@article{Watson1953,
  title   = {Molecular structure of nucleic acids: a structure for deoxyribose nucleic acid},
  author  = {Watson, James D. and Crick, Francis H. C.},
  journal = {Nature},
  volume  = {171}, number = {4356}, pages = {737--738}, year = {1953}, doi = {10.1038/171737a0}
}

@article{Yakovchuk2006,
  title   = {Base-stacking and base-pairing contributions into thermal stability of the {DNA} double helix},
  author  = {Yakovchuk, Peter and Protozanova, Ekaterina and Frank-Kamenetskii, Maxim D.},
  journal = {Nucleic Acids Research},
  volume  = {34}, number = {2}, pages = {564--574}, year = {2006}, doi = {10.1093/nar/gkj454}
}

@article{Seifert2012,
  title   = {Stochastic thermodynamics, fluctuation theorems and molecular machines},
  author  = {Seifert, Udo},
  journal = {Reports on Progress in Physics},
  volume  = {75}, number = {12}, pages = {126001}, year = {2012}, doi = {10.1088/0034-4885/75/12/126001}
}

@article{Gaspard2004,
  title   = {Fluctuation theorem for nonequilibrium reactions},
  author  = {Gaspard, Pierre},
  journal = {Journal of Chemical Physics},
  volume  = {120}, number = {19}, pages = {8898--8905}, year = {2004}, doi = {10.1063/1.1688736}
}

@article{Rao2016,
  title   = {Nonequilibrium thermodynamics of chemical reaction networks: wisdom from stochastic thermodynamics},
  author  = {Rao, Riccardo and Esposito, Massimiliano},
  journal = {Physical Review X},
  volume  = {6}, number = {4}, pages = {041064}, year = {2016}, doi = {10.1103/PhysRevX.6.041064}
}

@article{Bell1978,
  title   = {Models for the specific adhesion of cells to cells},
  author  = {Bell, George I.},
  journal = {Science},
  volume  = {200}, number = {4342}, pages = {618--627}, year = {1978}, doi = {10.1126/science.347575}
}

@article{Eyring1935,
  title   = {The activated complex in chemical reactions},
  author  = {Eyring, Henry},
  journal = {Journal of Chemical Physics},
  volume  = {3}, number = {2}, pages = {107--115}, year = {1935}, doi = {10.1063/1.1749604}
}

@article{Klein2018,
  title   = {Hybridization kinetics explains {CRISPR-Cas} off-targeting rules},
  author  = {Klein, Misha and Eslami-Mossallam, Behrouz and Arroyo, Daniel G. and Depken, Martin},
  journal = {Cell Reports},
  volume  = {22}, pages = {1413--1423}, year = {2018}, doi = {10.1016/j.celrep.2018.01.045}
}

@article{EslamiMossallam2022,
  title   = {A kinetic model predicts {SpCas9} activity, improves off-target classification, and reveals the physical basis of targeting fidelity},
  author  = {Eslami-Mossallam, Behrouz and Klein, Misha and Smagt, Constantijn V. D. and Sanden, Koen V. D. and Jones, Stephen K. and Hawkins, John A. and Finkelstein, Ilya J. and Depken, Martin},
  journal = {Nature Communications},
  volume  = {13}, pages = {1367}, year = {2022}, doi = {10.1038/s41467-022-28994-2}
}

@article{Alkan2018,
  title   = {{CRISPR-Cas9} off-targeting assessment with nucleic acid duplex energy parameters},
  author  = {Alkan, Ferhat and Wenzel, Anne and Anthon, Christian and Havgaard, Jakob Hull and Gorodkin, Jan},
  journal = {Genome Biology},
  volume  = {19}, pages = {177}, year = {2018}, doi = {10.1186/s13059-018-1534-x}
}

@article{Hopfield1974,
  title   = {Kinetic proofreading: a new mechanism for reducing errors in biosynthetic processes requiring high specificity},
  author  = {Hopfield, J. J.},
  journal = {Proceedings of the National Academy of Sciences},
  volume  = {71}, number = {10}, pages = {4135--4139}, year = {1974}, doi = {10.1073/pnas.71.10.4135}
}

@article{SartoriPigolotti2015,
  title   = {Thermodynamics of error correction},
  author  = {Sartori, Pablo and Pigolotti, Simone},
  journal = {Physical Review X},
  volume  = {5}, number = {4}, pages = {041039}, year = {2015}, doi = {10.1103/PhysRevX.5.041039}
}

@article{Chen2017,
  title   = {Enhanced proofreading governs {CRISPR-Cas9} targeting accuracy},
  author  = {Chen, Janice S. and Dagdas, Yavuz S. and Kleinstiver, Benjamin P. and Welch, Moira M. and Sousa, Alexander A. and Harrington, Lucas B. and Sternberg, Samuel H. and Joung, J. Keith and Yildiz, Ahmet and Doudna, Jennifer A.},
  journal = {Nature},
  volume  = {550}, number = {7676}, pages = {407--410}, year = {2017}, doi = {10.1038/nature24268}
}

\clearpage
\onecolumngrid

\begin{center}
{\Large Supplemental Material for ``Nonequilibrium Dynamics of R-loop Formation in the CRISPR-Cas9 System''}
\end{center}

\setcounter{section}{0}
\renewcommand{\thesection}{S\arabic{section}}
\setcounter{figure}{0}
\renewcommand{\thefigure}{S\arabic{figure}}
\setcounter{table}{0}
\renewcommand{\thetable}{S\arabic{table}}
\setcounter{equation}{0}
\renewcommand{\theequation}{S\arabic{equation}}
\makeatletter
\def\@hangfrom@section#1#2#3{\@hangfrom{#1#2}\MakeTextUppercase{#3}}
\def\@sectioncntformat#1{\@seccntformat{#1}}
\makeatother

\section{Data split, optimization, and parameter constraints}
\label{appS:optim}

The two libraries are fitted independently. The model is identical for both libraries. The optimizer and the epoch budget are identical as well. Only the training subset changes between the variants discussed below. The split is not a plain random split: each partition first assigns to the training set at least one variant carrying each mismatch type and each position, so that the training set covers the full parameter set, and then fills the training set to a fraction drawn uniformly between $60\%$ and $70\%$ of the library, with the remaining variants divided equally between the holdout and test sets. Parameters are optimized with Adam, at a learning rate of $10^{-3}$, a weight decay of $10^{-4}$, and a gradient-norm clip of $1$, for $90000$ epochs on the training set mean squared error, $\mathcal{L}=\mathrm{MSE}_{\mathrm{train}}$. The held-out subset does not enter the objective but drives a \texttt{ReduceLROnPlateau} scheduler, and the test set is evaluated once after training, so neither set enters the gradient. Structural constraints, namely $c>0$, $0<\alpha_2<\alpha_1$, and nonnegative position coefficients, are imposed by soft penalties with a hard projection after each step. Initialization follows the physics of the model, with $\gamma_p$ from the absolute Lasso coefficients of the training subset, $\delta\theta$ from $1.56^{\circ}$, $\alpha_1=\alpha_2=0.1$, $c=1.0$, and $b_{\mathrm{rd}}=-4.0$. In the representative partition used in the main text the training, holdout, and test sets contain $1101$, $302$, and $302$ samples for r0 and $1169$, $275$, and $275$ samples for r349.

\section{Partition to partition variability and exclusion of extreme variants}
\label{appS:partitions}

\begin{figure}[!htbp]
\centering
\includegraphics[width=0.95\linewidth]{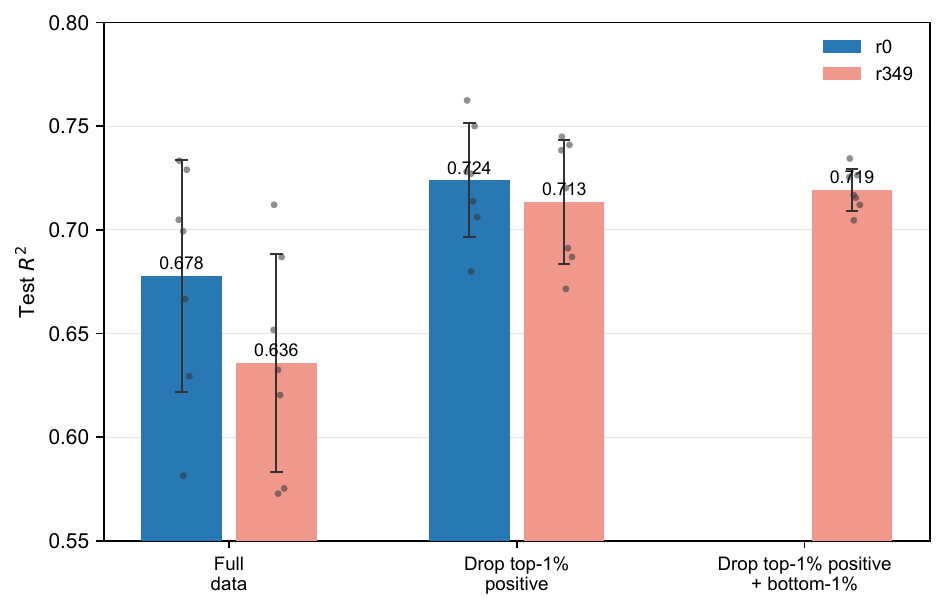}
\caption{Test $R^2$ for the three training variants, from left to right: the full data, the fit with the top-$1\%$ most active positive variants removed, and the fit with both the top-$1\%$ positive and the bottom-$1\%$ of the activity distribution removed. Bars give the mean over seven random partitions and error bars the standard deviation. Individual partitions are shown as dots. The third variant was run for r349 only, because for r0 the removal of the top-$1\%$ already restores Gaussian-like residuals (Fig.~\ref{fig:qq_after_r0}), so the further variants were not needed. The same model and optimizer are used throughout and only the training subset changes.}
\label{fig:r2_summary}
\end{figure}

On the held-out test sets the model reaches $R^2=0.68\pm0.06$ and $\mathrm{Spearman}=0.84\pm0.03$ for r0 and $R^2=0.64\pm0.05$ and $\mathrm{Spearman}=0.82\pm0.03$ for r349, averaged over seven independent random partitions (seeds $0$ to $6$). The rank correlations are insensitive to the choice of partition. Three training variants are compared in Fig.~\ref{fig:r2_summary}: the full data fit, the fit with the top-$1\%$ most active positive variants removed, and, for r349 only, the fit with both the top-$1\%$ positive and the bottom-$1\%$ of the activity distribution removed. For r0 the first variant was sufficient (Sec.~\ref{appS:residuals}), so the further variants were not run.

Removing the $18$ variants that are simultaneously positive and in the top $1\%$ of activity, about $1\%$ of each library, leaves $1687$ samples for r0 and $1701$ for r349. The identical model is refitted for the same seven partitions, the test set performance improves on both libraries, to $R^2=0.724\pm0.028$ for r0 and $0.713\pm0.030$ for r349 against $0.678\pm0.056$ and $0.636\pm0.053$ for the full data, and the partition to partition spread roughly halves.

\section{Residual diagnostics}
\label{appS:residuals}

\begin{figure}[!htbp]
	\centering
	\includegraphics[width=\textwidth]{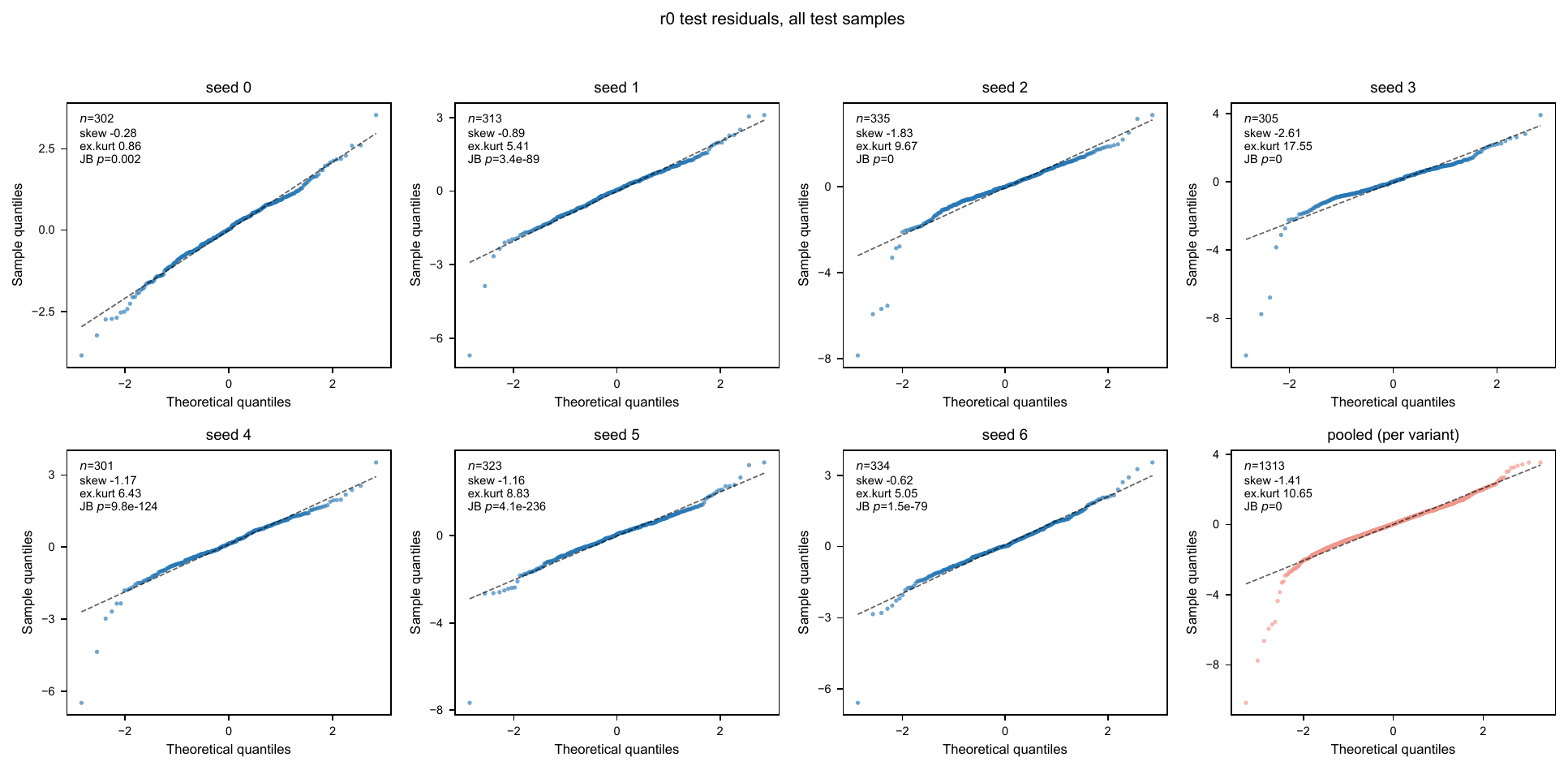}
	\caption{Q-Q plots of the r0 test set residuals, before removing any sample. Panels $1$--$7$ are the seven random partitions (seeds $0$--$6$). Panel $8$ pools them by assigning each variant a single residual, the mean over the partitions in which it falls in the test set, so that variants entering several test sets are counted once. A heavy left tail is present in every partition and is largest for seeds $2$--$4$. Each annotation gives the sample size $n$, the skewness, the excess kurtosis, and the Jarque--Bera $p$.}
	\label{fig:qq_before_r0}
\end{figure}

\begin{figure}[!htbp]
	\centering
	\includegraphics[width=\textwidth]{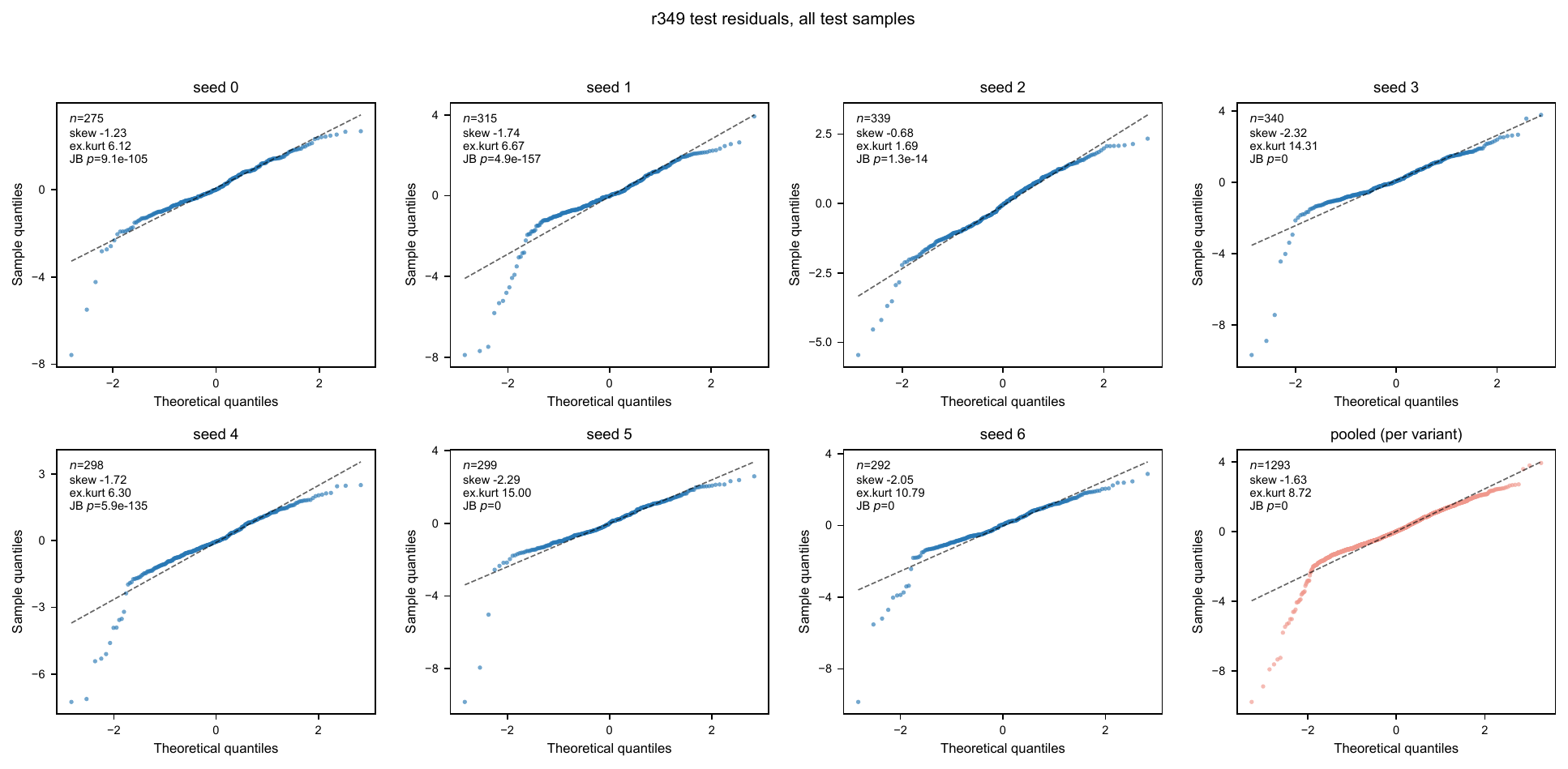}
	\caption{Same as Fig.~\ref{fig:qq_before_r0} but for the r349 test set residuals. The departure from Gaussianity is comparable in the two libraries, so neither library is singled out by the view across partitions.}
	\label{fig:qq_before_r349}
\end{figure}

\begin{figure}[!htbp]
	\centering
	\includegraphics[width=0.85\linewidth]{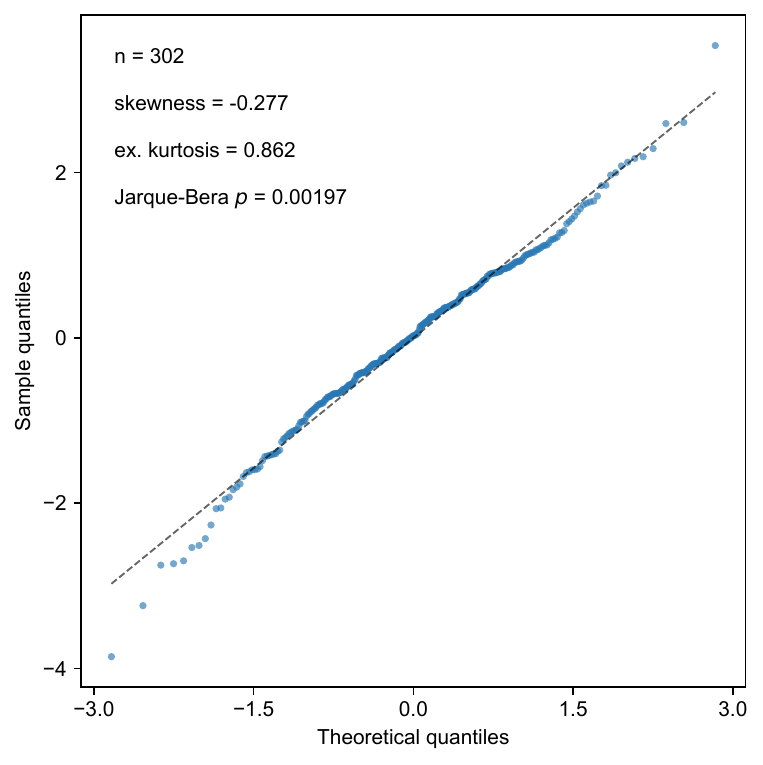}
	\caption{Q-Q plot of the r0 test set residuals ($n=302$).}
	\label{fig:qq_all}
\end{figure}

\begin{figure}[!htbp]
	\centering
	\includegraphics[width=0.85\linewidth]{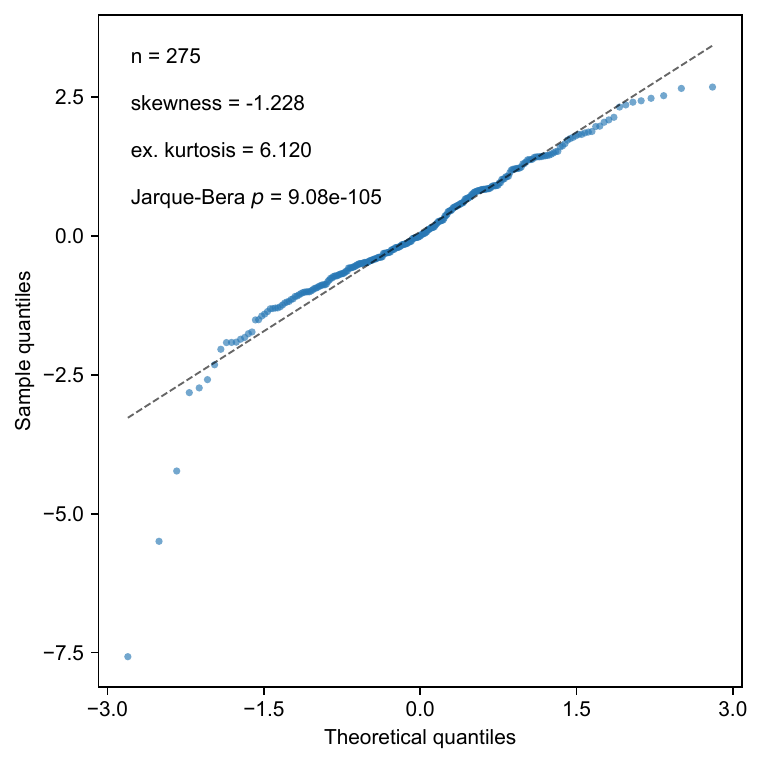}
	\caption{Q-Q plot of the r349 test set residuals ($n=275$).}
	\label{fig:qq_r349}
\end{figure}

Biological measurements carry noise, so we test whether the fit has absorbed the deterministic part of the data. We judge Gaussian-like noise by the first four moments of the residuals $r_i=\hat{y}_i-y_i$, since what is required of a fit is that the residuals carry no systematic offset, no skewness, and no excess kurtosis. We summarize the last two by the Jarque--Bera statistic and, for transparency, also report two stricter full-shape tests, a parametric-bootstrap Kolmogorov--Smirnov test with $9999$ resamples that re-estimates the mean and standard deviation each time, and the Shapiro--Wilk test, both of which are more readily rejected because they also probe shape features unrelated to skewness and kurtosis. Across the seven partitions the test residuals of both libraries carry a heavy left tail whose size fluctuates with the partition: the excess kurtosis spans $0.9$ to $17.6$ for r0 (median $6.4$) and $1.7$ to $15.0$ for r349 (median $6.7$), so Jarque--Bera rejects Gaussianity in every partition (Figs.~\ref{fig:qq_before_r0} and~\ref{fig:qq_before_r349}). The representative partition used in the main text, seed $0$, is the mildest case for r0 and is therefore not typical. It is retained only for the parameter discussion. Per-partition Q-Q plots of the two test sets on their own are shown in Figs.~\ref{fig:qq_all} and~\ref{fig:qq_r349}.

Removing the extreme residuals, a few percent of each test set, restores Gaussianity for r0 in every partition and for r349 in two of seven. The removed samples are almost exclusively double mismatches: pooling the seven partitions, $100\%$ of the r349 and $93\%$ of the r0 outliers carry two mismatches, whereas single mismatches are flagged in only one variant, and only in four of the seven partitions. The flagged variants also sit at the ends of the activity distribution: pooling the partitions, $88\%$ of the r0 and $90\%$ of the r349 outliers fall in the extreme quintiles of their test set's activity, dominated by the high-activity end ($63\%$ and $74\%$, respectively). Within the double mismatch class the deviation is therefore concentrated where the two mismatches lie close in sequence and where the measured activity is positive. For r349, $95\%$ of all positive-activity variants are flagged as outliers and their residuals are systematically negative, that is, the model underpredicts them. The residual cloud against measured activity shows this directly (Figs.~\ref{fig:resid_activity_r0} and~\ref{fig:resid_activity_r349}): it is centred on zero for $\mathrm{score}\le0$ and turns strongly negative for $\mathrm{score}>0$, with the flagged points lying in the positive-activity region.

\begin{figure}[!htbp]
	\centering
	\includegraphics[width=\textwidth]{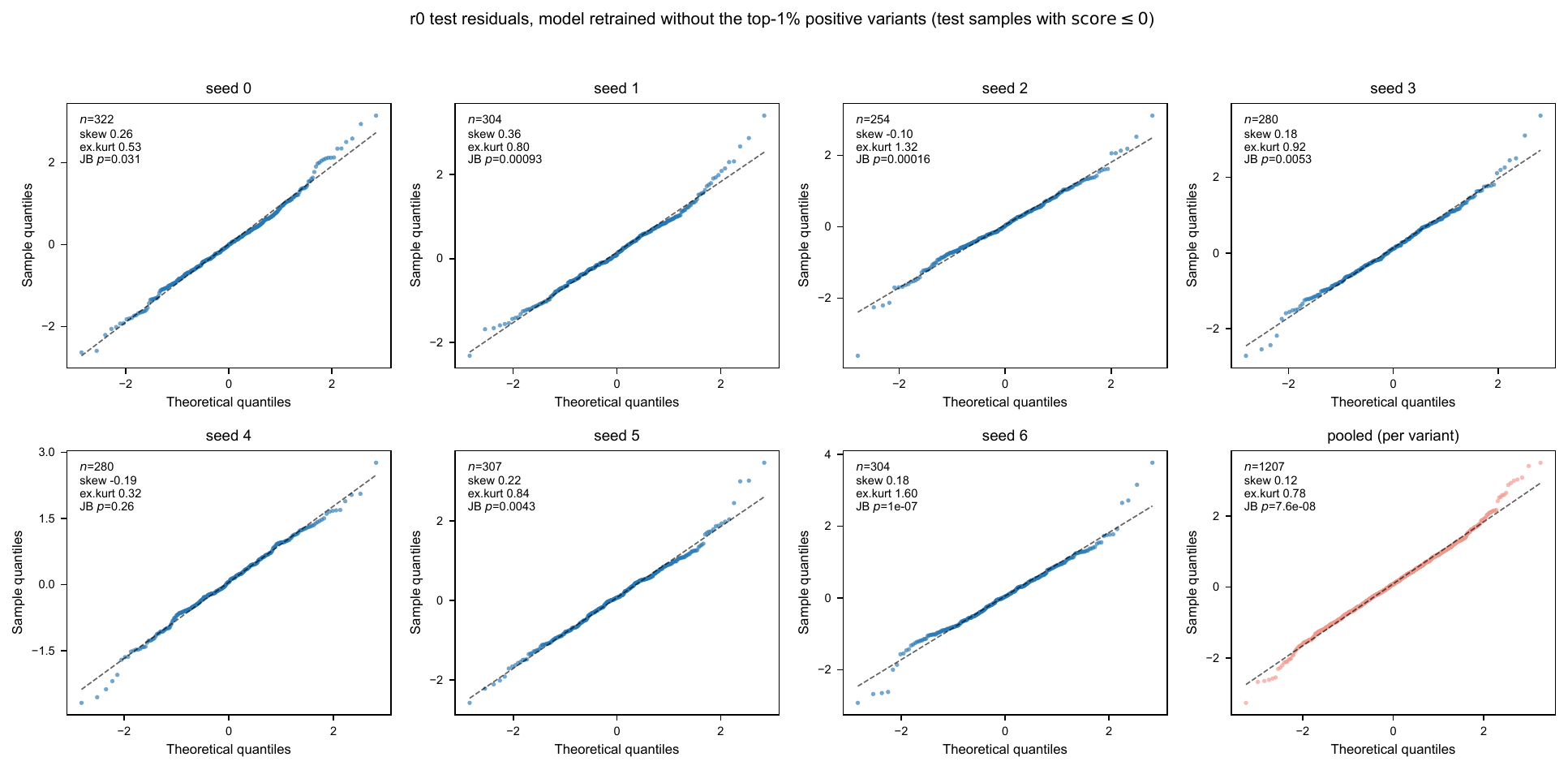}
	\caption{Q-Q plots of the r0 test residuals for the model \emph{retrained} after removing the $18$ variants that are both positive and in the top $1\%$ of activity ($1687$ samples per fit instead of $1705$). Layout as in Fig.~\ref{fig:qq_before_r0}, and only the test samples with $\mathrm{score}\le0$ are shown, the class that carries the heavy tail. With these variants out of the fit the excess kurtosis stays below $1.7$ in every partition while the test $R^2$ rises (mean $0.724\pm0.028$ against $0.678\pm0.056$ for the full data).}
	\label{fig:qq_after_r0}
\end{figure}

\begin{figure}[!htbp]
	\centering
	\includegraphics[width=\textwidth]{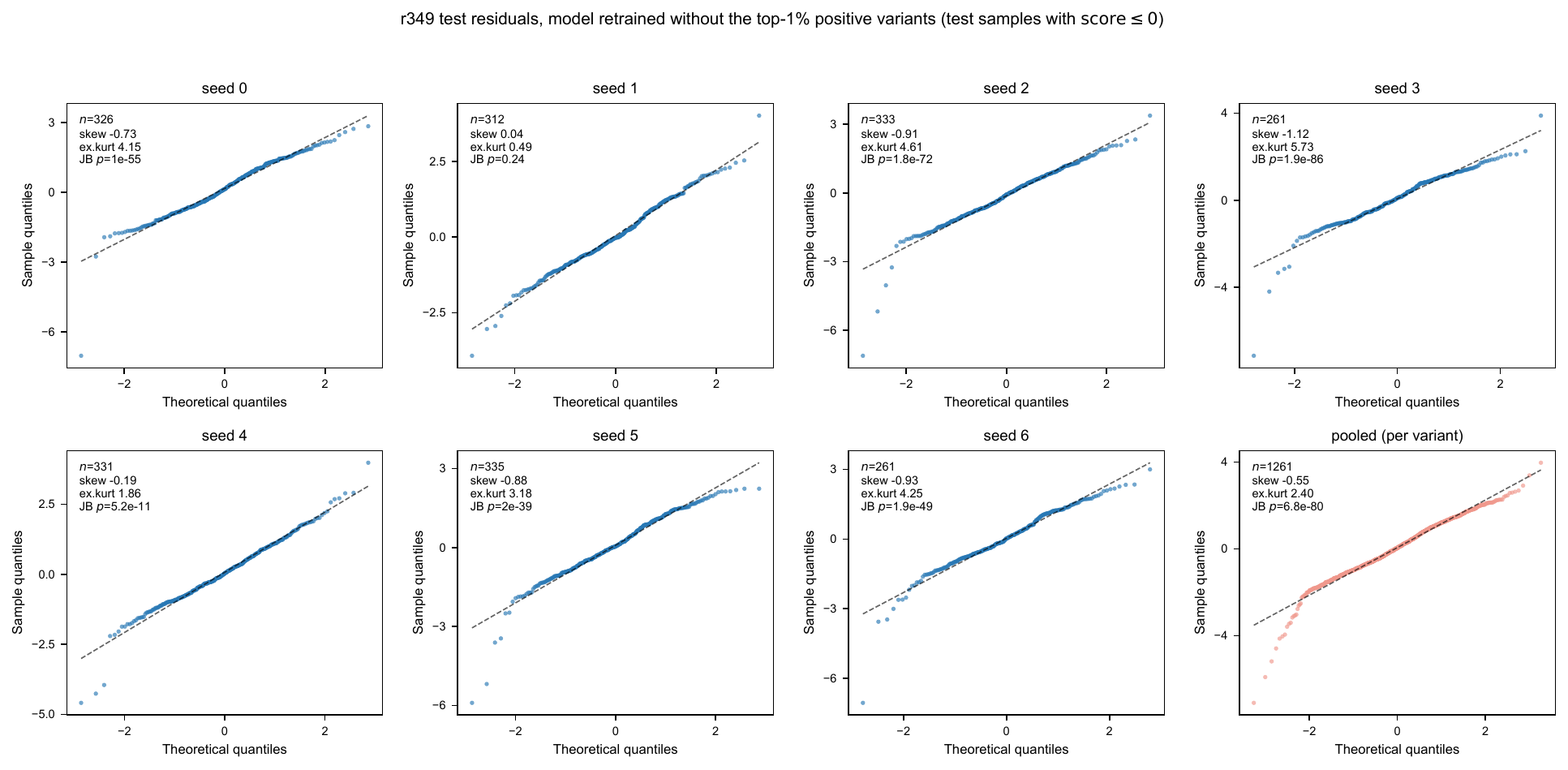}
	\caption{Same as Fig.~\ref{fig:qq_after_r0} but for r349 (the same $18$ variants removed, $1701$ samples per fit), again restricted to the test samples with $\mathrm{score}\le0$. Here the retraining raises the test $R^2$ from $0.636\pm0.053$ to $0.713\pm0.030$, but the excess kurtosis remains $3.5$ on average and a heavy left tail persists in six of the seven partitions, so for this library the departure is not carried by those few variants alone.}
	\label{fig:qq_after_r349}
\end{figure}

\begin{figure}[!htbp]
	\centering
	\includegraphics[width=\textwidth]{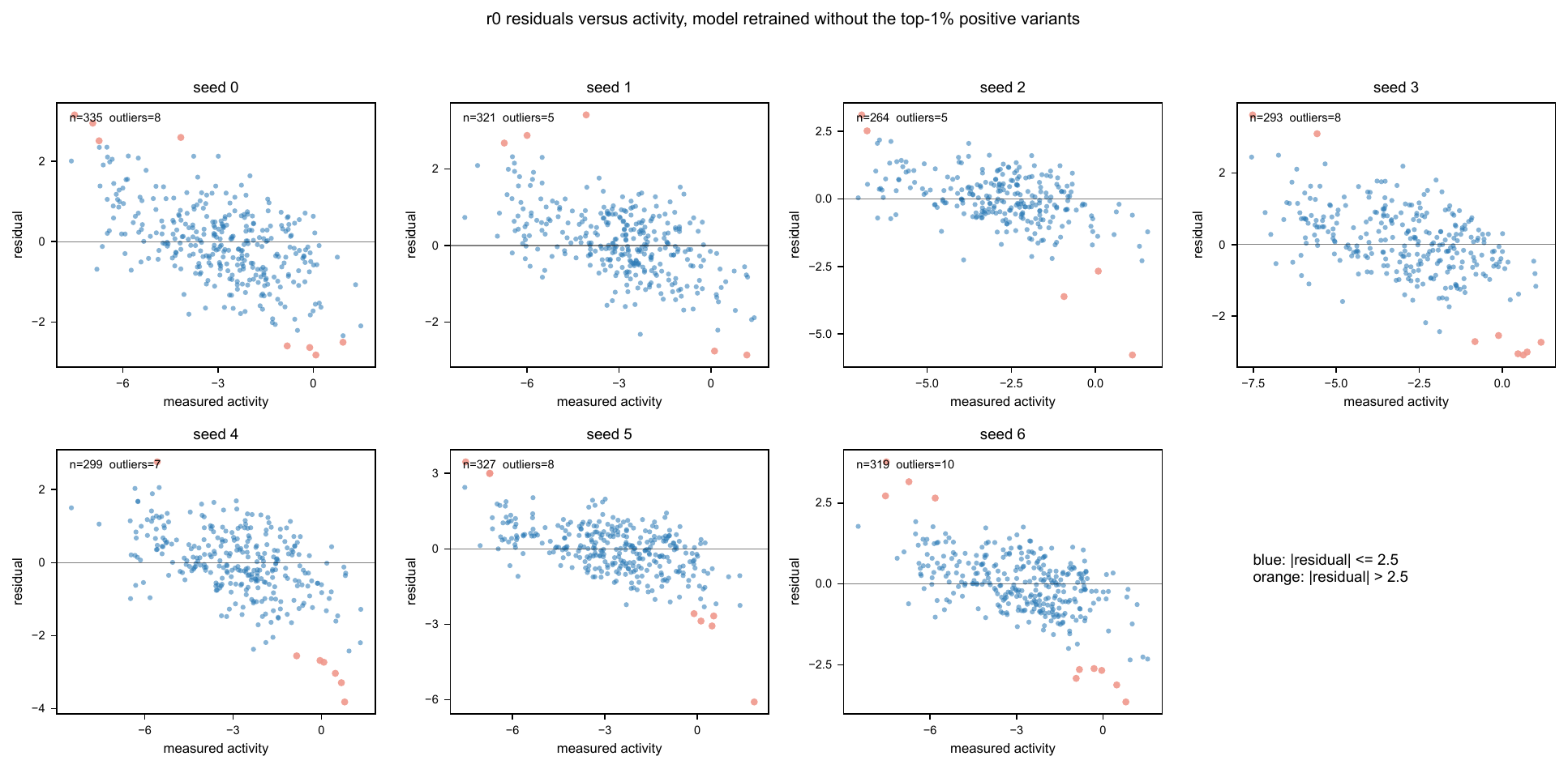}
	\caption{r0 residual against measured activity for the refitted model, one panel per partition. Points with $|r|>2.5$ are highlighted. The residual cloud is centred on zero for nonpositive activity and turns negative for positive activity, where the highlighted samples concentrate.}
	\label{fig:resid_activity_r0}
\end{figure}

\begin{figure}[!htbp]
	\centering
	\includegraphics[width=\textwidth]{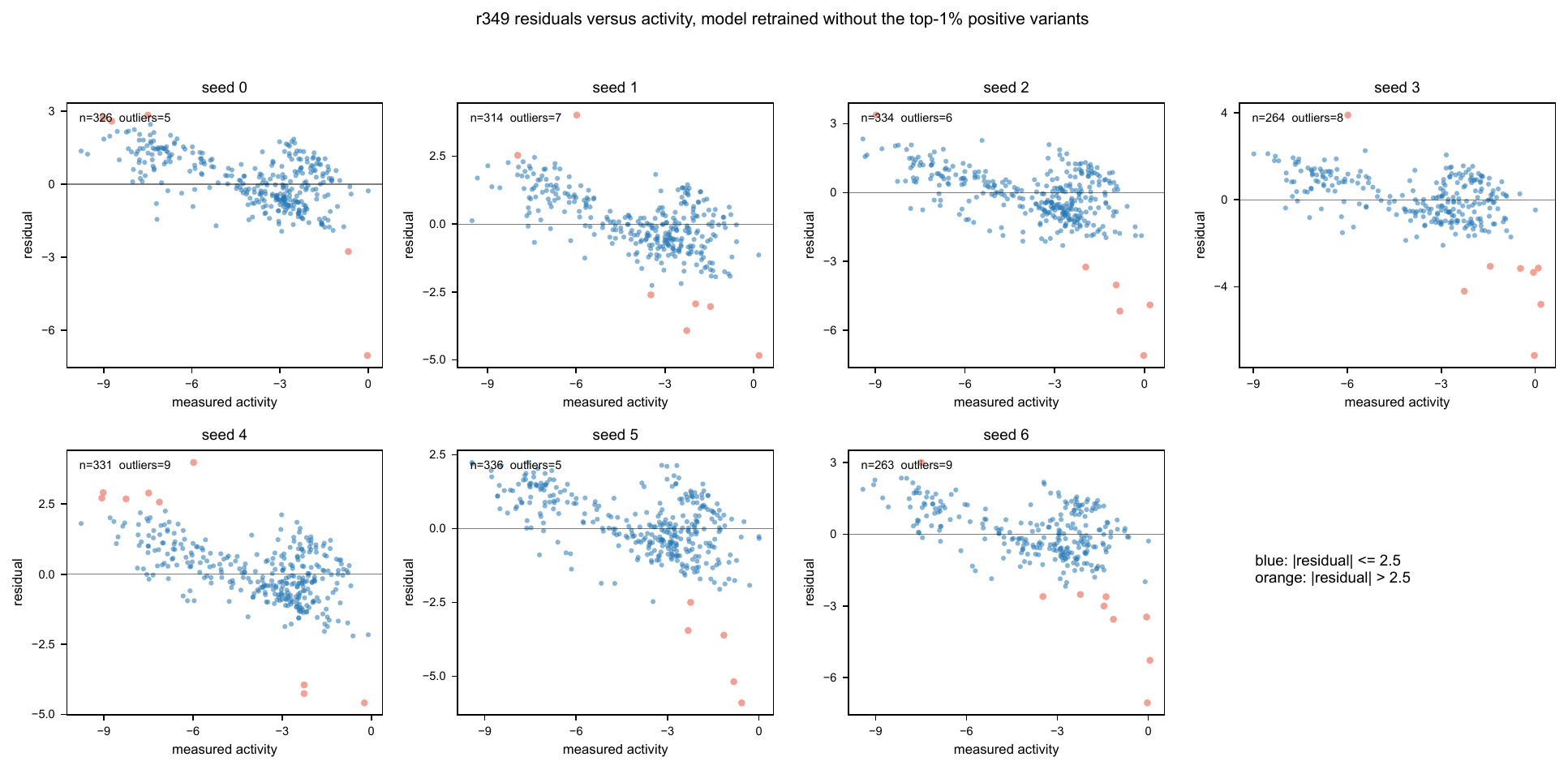}
	\caption{Same as Fig.~\ref{fig:resid_activity_r0} but for r349. Although the top-$1\%$ positive variants are no longer fitted, underpredictive double mismatch variants remain in every partition, which is why the r349 residuals stay heavy-tailed.}
	\label{fig:resid_activity_r349}
\end{figure}

Three origins of this departure should be kept distinct, and the present data separate them only partially. First, the samples are adversarial: two nearby mismatches are not independent, and a model that multiplies per position efficiencies, equivalently that adds per position free energies, cannot represent the synergy that lets such a variant retain activity, the coefficients $\alpha_k$ absorbing only part of it. Second, the limitation may lie in the model rather than in the data, because binding is a multistep reversible process whose outcome need not be a monotone function of an equilibrium free energy~\cite{Sternberg2014,Szczelkun2014}. Mismatches could also relax local strain, and the Arg71--guanine contact in Cas9~\cite{Anders2014} may soften the cost of dDrG mismatches together with the high stability of G-containing noncanonical pairs~\cite{Allawi1997}. The high-activity tail is then precisely where the free-energy mapping is incomplete. Third, experimental noise contributes, since counting fluctuations of low abundance variants and the nonlinear growth response inflate individual activities~\cite{Feng2021}, and that inflation is directed upward for rare active variants, the same direction required to produce the observed left tail. A mean near zero does not by itself identify Gaussian white noise, because the residual distribution mixes model bias, measurement fluctuation, and possible heteroscedasticity, so without replicate experiments the three contributions cannot be separated quantitatively.

To test whether a few extreme variants are responsible, we remove them and refit. The $18$ variants that are simultaneously positive and in the top $1\%$ of activity are dropped, leaving $1687$ samples for r0 and $1701$ for r349, and the identical model is refitted for the same seven partitions. The test set performance improves on both libraries, and the partition to partition spread roughly halves (Figs.~\ref{fig:qq_after_r0} and~\ref{fig:qq_after_r349}). The residuals of the refitted model are then Gaussian-like for r0 (excess kurtosis $0.89$ over the test samples with $\mathrm{score}\leq0$) but still heavy-tailed for r349 (excess kurtosis $3.51$), and the residual-versus-activity panels show why (Figs.~\ref{fig:resid_activity_r0} and~\ref{fig:resid_activity_r349}): for r0 the flagged points are gone, whereas for r349 underpredictive double mismatch variants remain across the activity range. The heavy tail is therefore only partly attributable to the most extreme variants. For r349 it also reflects the double mismatch coupling that the product form cannot absorb.

Two further refits isolate the origin. First, removing \emph{all} positive-activity variants, not only the top $1\%$, and refitting r349 on the $\mathrm{score}\le0$ subset ($7$ partitions, $R^2=0.729\pm0.040$) leaves the tail in place: only two of the seven partitions pass the Jarque--Bera test (excess kurtosis $0.39$ and $-0.41$), the remaining five spanning $0.99$ to $9.86$. The flagged variants are again all double mismatches, now confined to the negative-activity range, and their residuals split by sign: variants with $\mathrm{score}<-3$ are overpredicted (mean $+2.8$ to $+4.0$) while those with $-3<\mathrm{score}<0$ are underpredicted (mean $-3.6$ to $-4.5$). This is the signature of regression dilution, the product form compressing the double mismatch activities toward the centre from both ends, so removing the upper tail alone cannot remove the heavy tail.

Second, removing in addition the bottom $1\%$ of the activity distribution, so that only central negative variants and weakly positive ones remain, and refitting r349 again ($7$ partitions, $R^2=0.719\pm0.010$, the most reproducible of the three variants) still leaves six of seven partitions non-Gaussian (excess kurtosis $0.75$ to $3.66$, and the pooled value falls from $3.01$ to $1.60$). The heavy tail therefore survives the removal of both extremes. This points to the model rather than to a few samples, the strongly active and strongly repressed variants being those for which a per position product cannot reproduce the measured activity, although a larger experimental error on such variants cannot be excluded from the present data. By contrast, r0 needs only the top-$1\%$ removal to become Gaussian-like: its positive-activity variants include single mismatches that the model handles, and only the extreme ones are mispredicted, whereas every positive-activity variant of r349 is a double mismatch and is strongly underpredicted.

\begin{figure}[!htbp]
	\centering
	\includegraphics[width=0.85\linewidth]{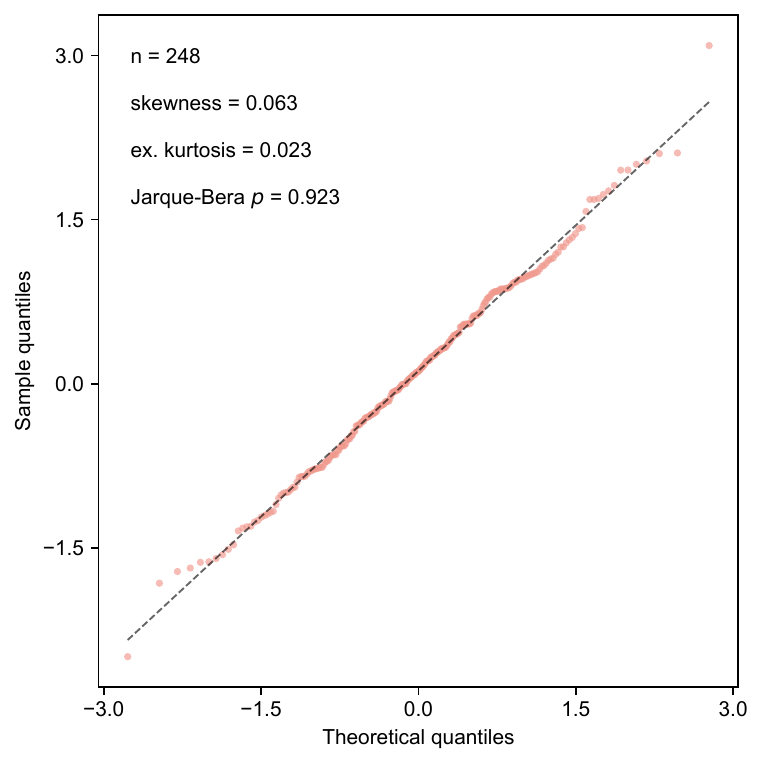}
	\caption{Q-Q plot of the r0 test set residuals for the model retrained on the negative-activity subset ($\mathrm{score}\le0$, $n=248$). Compare Fig.~\ref{fig:qq_all}.}
	\label{fig:qq_r0_nonpos}
\end{figure}

\begin{figure}[!htbp]
	\centering
	\includegraphics[width=0.85\linewidth]{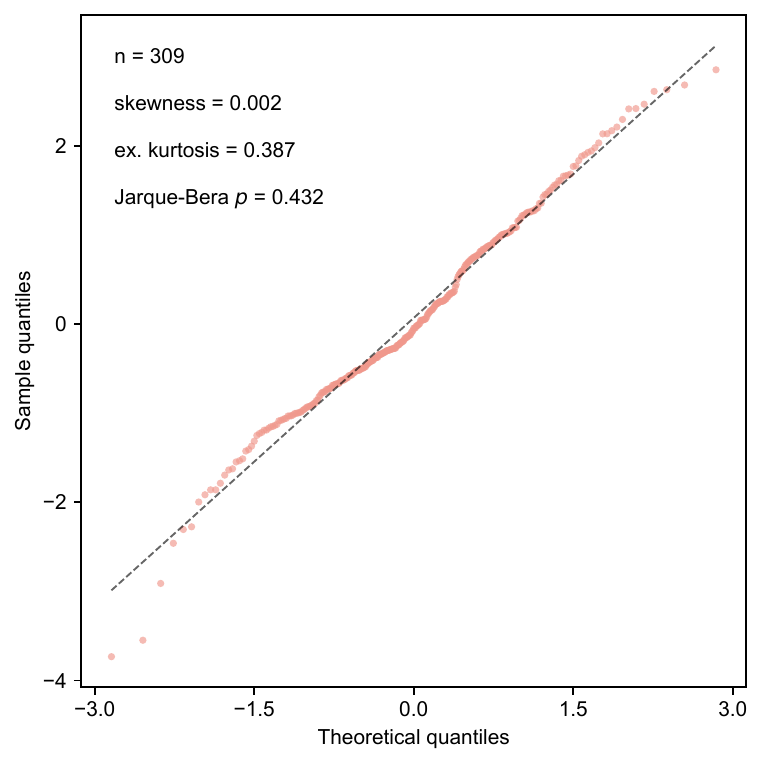}
	\caption{Q-Q plot of the r349 test set residuals for the model retrained on the negative-activity subset ($\mathrm{score}\le0$, $n=309$). Compare Fig.~\ref{fig:qq_r349}.}
	\label{fig:qq_r349_nonpos}
\end{figure}

\begin{figure}[!htbp]
	\centering
	\includegraphics[width=0.85\linewidth]{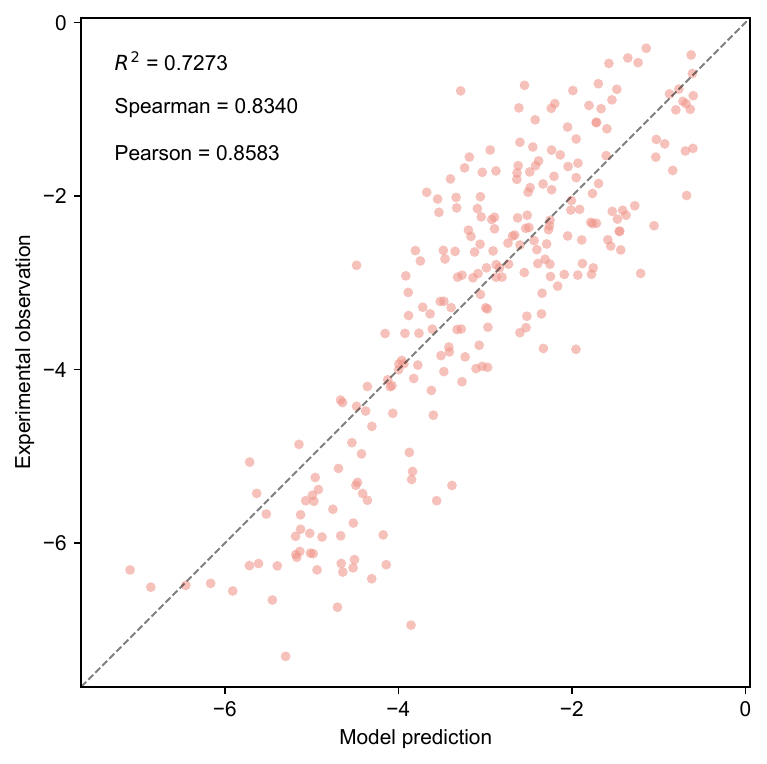}
	\caption{Predicted against measured activity for the r0 model retrained on the negative-activity subset ($\mathrm{score}\le0$). Test set, $n=248$.}
	\label{fig:prediction_r0_nonpos}
\end{figure}

\begin{figure}[!htbp]
	\centering
	\includegraphics[width=0.85\linewidth]{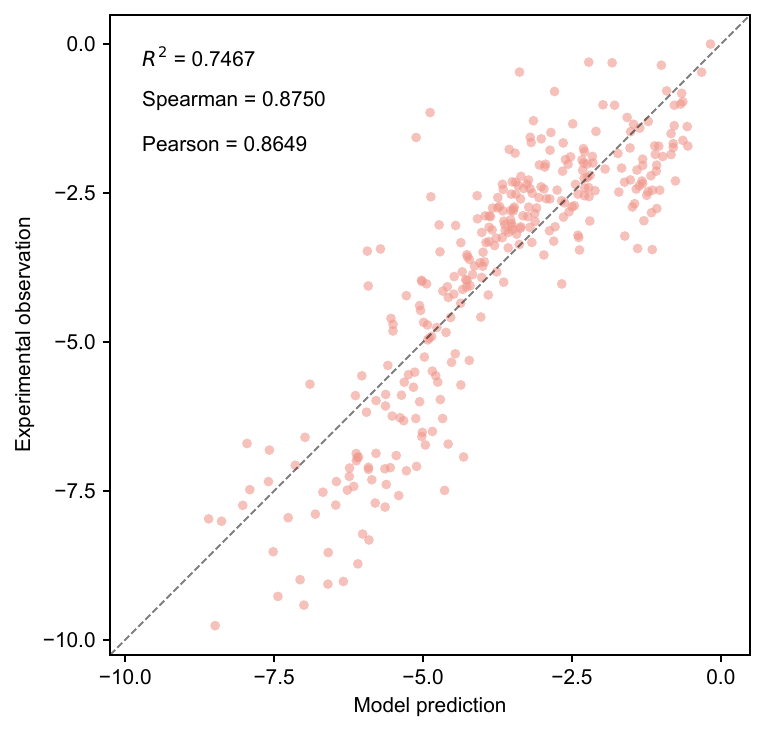}
	\caption{Predicted against measured activity for the r349 model retrained on the negative-activity subset ($\mathrm{score}\le0$). Test set, $n=309$.}
	\label{fig:prediction_r349_nonpos}
\end{figure}

We therefore test the model-limitation reading directly, by fitting the negative-activity bulk in isolation. Retraining the identical model on the subset with $\mathrm{score}\le0$ ($1597$ samples for r0 and $1696$ for r349, same optimizer and epoch budget) removes the tail: on the corresponding test sets r0 reaches $R^2=0.727$ with skewness $0.063$, excess kurtosis $0.023$, $\mathrm{JB}\ p=0.923$, KS $p=0.800$ and Shapiro $p=0.840$ ($n=248$), and r349 reaches $R^2=0.747$ with skewness $0.002$, excess kurtosis $0.387$, $\mathrm{JB}\ p=0.432$, KS $p=0.011$ and Shapiro $p=0.009$ ($n=309$) (Figs.~\ref{fig:qq_r0_nonpos}, \ref{fig:qq_r349_nonpos}, \ref{fig:prediction_r0_nonpos} and~\ref{fig:prediction_r349_nonpos}). The moment criterion is then met by both libraries, while the stricter full-shape tests still flag a marginal shape deviation in r349. Gaussian-like noise therefore describes the negative-activity bulk, which holds $94\%$ of the r0 and $99\%$ of the r349 variants, and the positive-activity, double mismatch tail is the regime in which the present free-energy mapping is incomplete.

\end{document}